\documentclass{mn}
\usepackage{amsfonts}
\usepackage{amsmath}
\usepackage{graphicx}
\def\gsim{\;\lower4pt\hbox{${\buildrel\displaystyle >\over\sim}$}\;}
\def\lsim{\;\lower4pt\hbox{${\buildrel\displaystyle <\over\sim}$}\;}
\def\grls{\;\lower4pt\hbox{${\buildrel\displaystyle >\over <}$}\;}
\begin{document}

\title[MHD stability criteria for composite MSIDs]
{Axisymmetric stability criteria for a composite system of
stellar and magnetized gaseous singular isothermal discs}
\author[Y.-Q. Lou and Y. Zou]{Yu-Qing Lou$^{1,2,3,4}$
and Yue Zou$^1$\\$^1$Physics Department and Tsinghua Center for
Astrophysics (THCA), Tsinghua University, Beijing 100084, China;
\\$^2$Centre de Physique des Particules de Marseille (CPPM)
/Centre National de la Recherche Scientifique (CNRS)\\
\qquad\quad
/Institut National de Physique Nucl\'eaire et de Physique
des Particules (IN2P3) et Universit\'e \\ \qquad\ \
de la M\'editerran\'ee Aix-Marseille II, 163, Avenue de 
Luminy Case 902 F-13288 Marseille, Cedex 09, France;
\\$^3$Department of Astronomy and Astrophysics, The University
of Chicago, 5640 Ellis Ave, Chicago, IL 60637, USA;
\\$^4$National Astronomical Observatories, Chinese Academy
of Science, A20, Datun Road, Beijing 100012, China.
\\}
\date{Accepted 2004... Received 2004...; in
original form 2004}\maketitle
%May 31, 2004 (Monday)        THCA noon.
%June 1, 2004 (Tuesday)       THCA morning.
%June 3, 2004 (Thursday)      Home evening.
%June 4, 2004 (Friday)        Home morning. THCA
%June 5, 2004 (Saturday)      Home morning.
%June 6, 2004 (Sunday)        Home early morning.
%July 1, 2004 (Thursday)      THCA morning.
%July 2, 2004 (Friday)        Home early morning; THCA
%January   6, 2005 (Thursday)    referee report for ME753; THCA
%February  7, 2005 (Monday)      THCA evening
%February 14, 2005 (Monday)      THCA night; Valentine's Day 
%February 16, 2005 (Wednesday)   THCA afternoon 
%February 17, 2005 (Thursday)    THCA morning, afternoon, evening and night
%February 18, 2005 (Friday)      THCA morning, afternoon, night
%February 19, 2005 (Saturday)    THCA morning, afternoon and night 
%February 20, 2005 (Sunday)      THCA early morning
%March 11,    2005 (Friday)      TIARA evening
%March 12,    2005 (Saturday)    Berkeley and TIARA morning, afternoon, night
%March 13,    2005 (Sunday)      Berkeley and TIARA morning, evening
%March 21,    2005 (Monday)      THCA afternoon 
%March 26,    2005 (Saturday)    THCA afternoon, evening 
%March 27,    2005 (Sunday)      THCA morning, afternoon, evening 
%August 3,    2005 (Wednesday)   CPPM/IN2P3/Universite Marseille, France
%                                 2nd report of the referee
%November  9,  2005 (Wednesday)  THCA morning, afternoon, evening
%November 10,  2005 (Thursday)   THCA night ME753 accepted 

\begin{abstract}
Using the fluid-magnetofluid formalism, we obtain axisymmetric 
stability criteria for a composite disc system consisting 
of stellar and gaseous magnetized singular isothermal discs 
(MSIDs). Both (M)SIDs are presumed to be razor-thin and are 
gravitationally coupled in a self-consistent axisymmetric 
background equilibrium with power-law surface mass densities 
and flat rotation curves. The gaseous MSID is embedded with a 
non-force-free coplanar azimuthal magnetic field $B_{\theta}(r)$ 
of radial scaling $r^{-1/2}$. Lou \& Zou recently
%(2004) 
constructed exact global stationary configurations for both axisymmetric 
and nonaxisymmetric coplanar magnetohydrodynamic (MHD) perturbations in 
such a composite MSID system and proposed the MHD $D_s-$criteria for 
axisymmetric stability by the hydrodynamic analogy. In a different 
perspective, we derive and analyze here the time-dependent WKBJ 
dispersion relation in the low-frequency and tight-winding regime to 
examine axisymmetric stability properties. By introducing a rotational 
Mach number $D_s$ for the ratio of the stellar rotation speed $V_s$ to 
the stellar velocity dispersion $a_s$, one readily determines the stable 
range of $D_s^2$ numerically to establish the $D_s-$criteria for 
axisymmetric MSID stability. Those MSID systems rotating either too fast 
(ring fragmentation) or too slow (Jeans collapse) are unstable. The stable 
range of $D_s^2$ depends on three dimensionless parameters: the ratio 
$\lambda$ for the Alfv\'en speed to the sound speed in the gaseous MSID, 
the ratio $\beta$ for the square of the stellar velocity dispersion to 
the gas sound speed and the ratio $\delta$ for the surface mass densities 
of the two (M)SIDs. Our WKBJ results of (M)SID instability provide 
physically compelling explanations for the stationary analysis of Lou 
\& Zou.
% (2004). 
We further introduce an effective MHD $Q$ parameter for a composite 
MSID system and compare with the earlier work of Elmegreen,
%(1995), 
Jog
%(1996) 
and Shen \& Lou.
%(2003, 2004a, b).  
As expected, an axisymmetric dark matter halo enhances the stability 
against axisymmetric disturbances in a composite partial MSID system.
In terms of the global star formation rate in a disc galaxy system, 
it would appear physically more sensible to examine the MHD $Q_M$ 
stability criterion against galactic observations. Relevance 
to large-scale structures in disc galaxies are also discussed.
\end{abstract}


\section{INTRODUCTION}

In contexts of galactic structures, disc stabilities and global star 
formation rates in spiral galaxies, we derive here instability criteria 
for axisymmetric coplanar magnetohydrodynamic (MHD) perturbations in a 
composite disc system with an azimuthal magnetic field in the gas disc 
component and establish a generalized definition of an effective MHD 
$Q_M$ parameter appropriate to such a 
magnetized gravitational system. Formulated as such, this is an idealized 
and limited theoretical MHD disc problem yet with several key conceptual 
elements included. The simple physical rationale is that zonal regions of 
higher gas density and magnetic field are vulnerable to active formation 
of massive stars with various scales involved differing by many orders of 
magnitudes. On the same ground, non-axisymmetric stability criteria are 
equally important but are more challenging to establish (see e.g. Shu et 
al. 2000 for relevant issues). Over four decades, important development 
have been made for instability criteria relevant to galactic disc dynamics 
(see Lin 1987, Binney \& Tremaine 1987 and Bertin \& Lin 1996 and 
extensive references therein). The original studies of axisymmetric
instabilities were conducted by Safronov (1960) and Toomre (1964) who 
introduced the dimensionless $Q$ parameter for the local stability (i.e. 
$Q>1$) against axisymmetric ring-like disturbances. For disc galaxies, it 
would be more realistic and sensible to investigate a composite disc system 
consisting of a stellar disc, a magnetized gas disc and a massive dark 
matter halo. There have been extensive theoretical studies on this type 
of composite two-component disc systems of various sub-combinations (Lin 
\& Shu 1966, 1968; Kato 1972; Jog \& Solomon 1984a; Bertin \& Romeo 1988; 
Romeo 1992; Elmegreen 1995; Jog 1996; Lou \& Fan 1998b, 2000a, b; Lou \& 
Shen 2003; Shen \& Lou 2003, 2004a, b; Lou \& Zou 2004; Lou \& Wu 2005). 
In particular, there have been several studies trying to define a proper 
effective $Q$ parameter for local axisymmetric instability criterion in a 
composite disc system (Elmegreen 1995; Jog 1996; Lou \& Fan 1998b, 2000a, b; 
Shen \& Lou 2003, 2004a, b). From different perspectives, these analyses 
offer insights for instability properties of a composite disc system and 
provide a theoretical basis for understanding the large-scale dynamics in 
such a system (e.g. Lou \& Fan 2000a, b; Lou \& Shen 2003; Shen \& Lou 
2003, 2004a, b; Lou \& Zou 2004; Lou \& Wu 2005).

The main motivation here is to explore basic properties the MSID 
model in a composite system and obtain conceptual insights for
astrophysical applications in magnetized spiral galaxies and in 
estimating global star formation rates in disc galaxies (e.g.,
Lou \& Bian 2005). In theoretical studies of modeling disc galaxies, 
the class of SID models has a distinguished history since the pioneering 
work of Mestel (1963) (Zang 1976; Toomre 1977; Lemos, Kalnajs \& 
Lynden-Bell 1991; Lynden-Bell \& Lemos 1993; Syer \& Tremaine 
1996; Goodman \& Evans 1999; Charkrabarti, Laughlin \& Shu 2003) 
and has gained considerable attention and interests recently by 
considering a composite disc system and by incorporating effects 
of magnetic field (Shu et al. 2000; Lou 2002; Lou \& Fan 2002; Lou 
\& Shen 2003; Shen \& Lou 2003, 2004a, b; Shen, Liu \& Lou 2004; 
Lou \& Zou 2004; Lou \& Wu 2005). Specifically, Shu et al. (2000; 
see also Galli et al. 2001) studied global stationary (i.e., zero 
pattern speed) perturbation configurations in an isopedically 
magnetized SID without invoking the usual WKBJ or tight-winding 
approximation. They obtained exact global solutions for both aligned 
and unaligned axisymmetric and non-axisymmetric logarithmic spiral 
configurations and interpreted the axisymmetric solution for 
perturbations with radial propagations as demarcating the boundaries 
between the stable and unstable regimes. By these axisymmetric 
instabilities, a SID with a sufficiently slow rotation speed would
Jeans collapse induced by perturbations of larger radial scales, 
while a SID with a sufficiently fast rotation speed may suffer the 
ring fragmentation instability induced by perturbations of smaller 
radial scales (see fig. 2 of Shu et al. 2000). By introducing a 
rotational Mach number $D$, defined as the ratio of the SID rotation 
speed $V$ to the isothermal sound speed $a$, the critical values of 
the highest and lowest $D$ for an axisymmetric stability can be 
determined directly from the marginal stability curve. To support 
their physical interpretations, they invoked the well-known Toomre 
$Q$ parameter and found that the highest $D$, namely the minimum of 
the ring fragmentation curve, corresponds to a $Q$ value very close 
to unity, thus heuristically suggesting the correspondence between 
the $D-$criterion and the $Q-$criterion.

Different from yet complementary to the analysis of Shu et al. (2000) on 
a single isopedically magnetized disc, Lou (2002) studied global coplanar 
MHD perturbations in a single MSID embedded with an azimuthal magnetic 
field and revealed that the minimum of the MHD ring fragmentation curve 
in this MSID model is tightly associated with the generalized MHD $Q_M$ 
parameter originally introduced by Lou \& Fan (1998a) in developing the 
galactic MHD density wave theory (Fan \& Lou 1996). For a composite 
system of two coupled unmagnetized hydrodynamic SIDs, Lou \& Shen (2003) 
constructed stationary global perturbation configurations and Shen \& Lou 
(2003) suggested a straightforward $D-$criterion for the axisymmetric ring 
fragmentation instability in such a system on the basis of a low-frequency
WKBJ analysis; they revealed that the minimum of the ring fragmentation in 
their composite SID model is again closely related to a 
proper effective $Q$ parameter (Elmegreen 1995; Jog 1996). Furthermore, 
for a composite SID system with an isopedically magnetized gaseous SID 
and a stellar SID in the fluid description, Lou \& Wu (2005) have 
constructed global stationary MHD perturbation structures and examined 
stability properties to anticipate a similar $D-$criterion in parallel 
to the case of Shen \& Lou (2003). Meanwhile, Shen \& Lou (2004b) have 
further generalized both work of Syer \& Tremaine (1996) and Lou \& Shen 
(2003) to the situation of a composite system for two gravitationally 
coupled scale-free discs; they also studied the axisymmetric stability
of such a composite system in terms of the marginal stability curves 
and proposed a $D_s-$criterion by the analogy of Shen \& Lou (2003).

We have recently examined two-dimensional coplanar MHD perturbations 
in a composite system consisting of a stellar SID and a gaseous MSID. 
Both SIDs are expediently approximated as razor-thin and the gas disc 
is embedded with a non-force-free coplanar magnetic field (Lou \& 
Zou 2004). In this fluid-magnetofluid MSID model approximation, we 
obtained exact global stationary MHD solutions for aligned and unaligned 
logarithmic spiral perturbation configurations in such a composite 
MSID system, expressed in terms of the stellar rotational Mach number 
$D_s$. In reference to the results of a single SID (Shu et al. 2000; 
Galli et al. 2001; Lou 2002), it would be natural to suggest that the 
stationary axisymmetric solutions with radial propagations give rise to 
marginal stability curves (see Figure \ref{f2} in this paper later and 
Lou \& Zou 2004). In comparison with the single SID case, the stable 
range of $D_s^2$ is reduced as a result of the mutual gravitational 
coupling between the two SIDs. However, in comparison with the case 
of a composite unmagnetized SID system (Lou \& Shen 2003; Shen \& 
Lou 2003), the stable range of $D_s^2$ expands considerably due to 
the presence of a coplanar magnetic field.

To confirm heuristic arguments for the above analogy and our 
intuitive physical interpretations, we conduct in this paper a 
low-frequency time-dependent stability analysis in the WKBJ or 
tight-winding approximation for the composite MSID system (Shen 
\& Lou 2003, 2004a; Lou \& Zou 2004). We shall demonstrate 
unambiguously the validity of demarcating the stable and unstable 
regimes by the stellar rotational Mach number $D_s$. To place our 
analysis in proper contexts, we also discuss specifically how the 
two effective $Q$ parameters of Elmegreen (1995) and of Jog (1996) 
are related to our $D_s$ parameter when other relevant parameters 
are specified, and show that the two effective $Q$ parameters are 
pertinent to the ring fragmentation instability in a composite 
MSID system.

In Section 2, we derive the time-dependent dispersion relation using 
the WKBJ approximation for perturbations in a composite MSID system 
and introduce a few key dimensionless parameters. In Section 3, we 
present the $D_s-$criterion and two effective $Q$ parameters for a 
composite MSID system being stable against arbitrary axisymmetric 
perturbations with radial propagations. Main results and discussions 
are summarized in Section 4.

\section{A FLUID-MAGNETOFLUID FORMALISM}

We consider below a composite system consisting of two gravitationally 
coupled (M)SIDs with one of the SIDs being magnetized and thus referred 
to as MSID. For physical variables, we use either superscript or 
subscript $s$ to indicate an association with the stellar SID and 
either superscript or subscript $g$ to indicate an association with 
the gaseous MSID. The stellar SID and the gaseous MSID can have 
constant yet different rotational speeds $V_s$ and $V_g$ (related to 
the phenomenon of asymmetric drift in the galactic context); we thus 
write the background angular rotation speeds $\Omega_s$ of the stellar 
SID and $\Omega_g$ of the gaseous MSID as
\begin{equation}
\Omega_s=V_s/r=a_sD_s/r
\end{equation}
and
\begin{equation}
\Omega_g=V_g/r=a_gD_g/r\ ,
\end{equation}
separately, where $a_s$ and $a_g$ are the constant velocity dispersion 
of the stellar SID and the isothermal sound speed of the gaseous MSID, 
respectively; $D_s$ and $D_g$ are the corresponding rotational Mach 
numbers. The relevant epicyclic frequencies in terms of $\Omega_s$ 
and $\Omega_g$ are given by 
\begin{equation}
\kappa^2_s\equiv
\frac{2\Omega_s}r\frac{d}{dr}(r^2\Omega_s)=2\Omega^2_s
\end{equation}
and
\begin{equation}
\kappa^2_g\equiv
\frac{2\Omega_g}r\frac{d}{dr}(r^2\Omega_g)=2\Omega^2_g\ ,
\end{equation}
respectively. Similar to a single MSID, we take the 
background azimuthal magnetic field to be in the form of
\begin{equation}\label{118}
B_\theta(r)={\cal{F}}r^{-1/2}\ ,
\end{equation}
where ${\cal{F}}$ is a constant
(Lou 2002; Lou \& Fan 2002) and
\begin{equation}\label{119}
B_r=B_z=0\ .
\end{equation}
For a more general power-law radial variation of the azimuthal 
magnetic field and those of other related background variables, 
the interested reader is referred to a recent analysis of Shen, 
Liu \& Lou (2005).

In the fluid approximation of a stellar SID, the mass conservation,
the radial component of the momentum equation and the azimuthal
component of the momentum equation are given below in order, namely
\begin{equation}\label{21}
\frac{\partial\Sigma^s}{\partial t}
+\frac{1}{r}\frac{\partial (r\Sigma^s u^s)}{\partial r}
+\frac1{r^2}\frac{\partial(\Sigma^sj^s)}{\partial\theta}=0\ ,
\end{equation}
\begin{equation}\label{22}
\frac{\partial u^s}{\partial t}+u^s\frac{\partial u^s}
{\partial r}+\frac{j^s}{r^2}\frac{\partial u^s}
{\partial\theta}-\frac{j^{s2}}{r^3}
=-\frac{1}{\Sigma^{s}}\frac{\partial\Pi^s}{\partial r}
-\frac{\partial\varphi}{\partial r}\ ,
\end{equation}
\begin{equation}\label{23}
\frac{\partial j^s}{\partial t}+u^s\frac{\partial j^s}
{\partial r}+\frac{j^s}{r^2}\frac{\partial j^s}
{\partial\theta}=-\frac{1}{\Sigma^{s}}\frac{\partial
\Pi^s}{\partial\theta}-\frac{\partial\varphi}{\partial\theta}\ ,
\end{equation}
where $u^s$ is the radial component of the bulk flow velocity of 
the stellar SID, $j^s\equiv rv^s$ is the stellar specific angular
momentum along the $\hat{z}$ direction, $v^s$ is the azimuthal
component of the stellar bulk flow velocity, $\varphi$ is the total 
gravitational potential, $\Pi^s$ is the vertically integrated pressure, 
and $\Sigma^s$ is the surface mass density of the stellar SID.

In the magnetofluid approximation for the gaseous MSID, the mass
conservation, the radial component of the momentum equation and
the azimuthal component of the momentum equation are given below
in order, namely
\begin{equation}\label{24}
\frac{\partial\Sigma^g}{\partial t}+\frac{1}{r} 
\frac{\partial (r\Sigma^g u^g)}{\partial r}
+\frac1{r^2}\frac{\partial(\Sigma^gj^g)}{\partial\theta}=0\ ,
\end{equation}
\begin{equation}\label{25}
\begin{split}
\frac{\partial u^g}{\partial t}+u^g\frac{\partial u^g}{\partial r}
+\frac{j^g}{r^2}\frac{\partial u^g}{\partial\theta}
-\frac{j^{g2}}{r^3}=-\frac1{\Sigma^g}
\frac{\partial\Pi^g}{\partial r}
-\frac{\partial\varphi}{\partial r} \quad   \\
\quad\qquad -\frac1{\Sigma^g}\int \frac{dzB_\theta}{4\pi r}
\bigg[\frac{\partial(rB_\theta)}{\partial r}
-\frac{\partial B_r}{\partial\theta}\bigg]\ ,
\end{split}
\end{equation}
\begin{equation}\label{26}
\begin{split}
\frac{\partial j^g}{\partial t}+u^g\frac{\partial j^g}{\partial r}
+\frac{j^g}{r^2}\frac{\partial j^g}{\partial\theta}
=-\frac1{\Sigma^g}\frac{\partial\Pi^g}{\partial\theta}
-\frac{\partial\varphi}{\partial\theta}\qquad\qquad \\
\qquad +\frac1{\Sigma^g}\int
\frac{dzB_r}{4\pi}\bigg[\frac{\partial(rB_\theta)}{\partial r}
-\frac{\partial B_r}{\partial\theta}\bigg]\ ,
\end{split}
\end{equation}
where $u^g$ is the radial component of the gas bulk flow velocity,
$j^g\equiv rv^g$ is the gas specific angular momentum in the
$\hat{z}$ direction, $v^g$ is the azimuthal component of the gas 
bulk flow velocity, $\Pi^g$ is the two-dimensional gas pressure,
$\Sigma^g$ is the gas surface mass density and $B_r$ and $B_\theta$ 
are the radial and azimuthal components of the magnetic field 
$\mathbf B$. The last two terms on the right-hand sides of equations 
(\ref{25}) and (\ref{26}) are the radial and azimuthal components 
of the Lorentz force due to the coplanar magnetic field. The two 
sets of fluid and magnetofluid equations $(\ref{21})-(\ref{23})$ 
and $(\ref{24})-(\ref{26})$ are dynamically coupled by the total 
gravitational potential $\varphi$ through the Poisson integral
\begin{equation}\label{27}
\varphi(r,\theta,t)=\oint d\psi{\int_0}^\infty
\frac{-G(\Sigma^g+\Sigma^s)\zeta d\zeta}
{[\zeta^2+r^2-2\zeta r\cos(\psi-\theta)]^{1/2}}\ .
\end{equation}
The gravitational effect of a massive dark matter halo is not 
included for the moment. The divergence-free condition for 
magnetic field ${\bf{B}}=(B_r, B_{\theta},\ 0)$ is simply 
\begin{equation}\label{28}
\frac{\partial(rB_r)}{\partial r}
+\frac{\partial B_\theta}{\partial\theta }=0\ ,
\end{equation}
and the radial and azimuthal components of the
magnetic induction equation for gas motions are
\begin{equation}\label{29}
\frac{\partial B_r}{\partial t}
=\frac{1}{r}\frac{\partial}{\partial\theta}(u^gB_\theta-v^gB_r)\ ,
\end{equation}
\begin{equation}\label{30}
\frac{\partial B_\theta}{\partial t}
=-\frac{\partial}{\partial r}(u^gB_\theta-v^gB_r)\ .
\end{equation}
Using Poisson integral ({\ref{27}), one readily
derives the following expressions for the
background surface mass densities
\begin{equation}\label{120}
\Sigma_0^s=\frac{a_s^2(1+D_s^2)}{2\pi Gr (1+\delta)}\
\end{equation}
and
\begin{equation}\label{121}
\Sigma_0^g=\frac{[a_g^2(1+D_g^2)-C_A^2/2]\delta}
{2\pi Gr (1+\delta)}\ ,
\end{equation}
where $\delta\equiv{\Sigma^g_0}/{\Sigma^s_0}$ is the surface mass 
density ratio of the two dynamically coupled background (M)SIDs and 
$C_A$ is the constant Alfv\'en wave speed in the MSID defined by
\begin{equation}
C_A^2\equiv\int dzB_\theta^2/(4\pi\Sigma^g_0)\ .
\end{equation}
Apparently, equations (\ref{120}) and (\ref{121}) requires
\begin{equation}\label{186}
a_s^2(1+D_s^2)=a_g^2(1+D_g^2)-C_A^2/2\ .
\end{equation}
We now introduce two more useful dimensionless parameters here. The 
first parameter $\beta\equiv a_s^2/a_g^2$ stands for the square of 
the ratio of the stellar velocity dispersion to the isothermal sound
speed in the MSID and the second parameter $\lambda^2\equiv C_A^2/a_g^2$ 
stands for the square of the ratio of the Alfv\'en wave speed to the 
isothermal sound speed in the MSID. In late-type disc galaxies, the 
stellar velocity dispersion $a_s$ is usually higher than the gas sound 
speed $a_g$, we thus focus on the case of $\beta\geq 1$ (Jog \& Solomon 
1984a, b; Bertin \& Romeo 1988; Jog 1996; Elmegreen 1995; Lou \& Fan 
1998b; Lou \& Shen 2003; Shen \& Lou 2003, 2004a, b; Lou \& Wu 2005).

Before going further, we note that a typical disc galaxy system involves 
a massive dark matter halo, a stellar disc and a gaseous disc of 
interstellar medium (ISM) on large scales, where the ISM disc is 
magnetized with the magnetic energy density ($\sim\hbox{1eV/cm}^3$) 
being comparable to the energy densities of thermal ISM and of 
relativistic cosmic-ray gas (e.g. Lou \& Fan 2003). To comprehensively 
understand multi-wavelength observations of large-scale spiral structures 
of disc galaxies and to develop potentially powerful observational 
diagnostics (Lou \& Fan 2000a, b), it would be more realistic and 
necessary to take into account of magnetic field effects in a composite 
magnetized disc-halo model.\footnote{The cosmic-ray gas is set aside 
here merely for the sake of simplicity. Relativistic cosmic-ray electrons 
gyrating around galactic magnetic field give off the observed synchrotron 
radiation).} While there are exceptions, galactic 
magnetic fields typically tend to be coplanar with the disc plane of 
a spiral galaxy on large scales. On smaller scales, regions of closed 
and open magnetic fields are most likely intermingled by the solar 
analogy (e.g. Lou \& Wu 2005). As a first step, Lou (2002) carried 
out a coplanar MHD perturbation analysis for stationary aligned and 
unaligned logarithmic spiral structures in a single MSID embedded 
with an azimuthal magnetic field and demonstrated that the minimum 
of the ring fragmentation curve in this MSID model is clearly related 
to the generalized MHD $Q_M$ parameter (Lou \& Fan 1998a). Since the 
background MHD rotational equilibrium adopted by Lou \& Fan (1998a) 
is not an MSID model, it would be more satisfying to justify the 
statement and interpretation of Lou (2002) in a dynamically 
self-consistent manner. We shall define a $Q_M$ parameter similar to 
that of Lou (2002) and show that this $Q_M$ is equivalent to that of 
Lou \& Fan (1998a).

For a single MSID, we readily derive linear equations (by setting 
relevant parameters for the stellar SID to vanish) for coplanar 
axisymmetric MHD perturbations with Fourier harmonic dependence 
$\exp(ikr+i\omega t)$, where $k$ is the radial wavenumber and 
$\omega$ is the angular frequency. In the usual WKBJ or tight-winding 
approximation of $kr\gg 1$, we obtain the local WKBJ dispersion 
relation for MHD density waves propagating in an MSID in the form of
\begin{equation}\label{1}
\omega^2=\kappa_g^2+k^2(a_g^2+C_A^2)-2\pi G|k|\Sigma^g_0\ ,
\end{equation}
which is the MHD generalization of the WKBJ dispersion 
relation for coplanar perturbations in an unmagnetized SID.

To derive an effective $Q_M$ parameter for the axisymmetric stability 
(i.e., $\omega^2\ge 0$) against MHD perturbations with an arbitrary 
$k$, the determinant of the right-hand side of equation (\ref{1}) 
should be negative for all $k$. This requires
\begin{equation}\label{12}
Q_M\equiv\frac{\kappa_g(C_A^2+a_g^2)^{1/2}}{\pi G\Sigma^g_0}>1\ ,
\end{equation}
where $Q_M$ is an generalized Toomre's $Q$ parameter for a single MSID. 
Inequality (\ref{12}) is of the same form as the effective $Q$ parameter 
derived by Lou \& Fan (1998a) for fast MHD density waves in a rigidly 
rotating gas disc with a different formalism but with the same assumption 
$kr\gg1$. This result establishes the connection found by Lou (2002) 
that the minimum of the ring fragmentation curve is tightly associated 
with the $Q_M$ parameter of Lou \& Fan (1998a).

In our low-frequency time-dependent analysis for a composite MSID 
system, we write coplanar axisymmetric MHD perturbations with the
same harmonic dependence of $\exp(ikr+i\omega t)$. In the WKBJ 
limit of $kr\gg 1$, we obtain the local WKBJ dispersion relation 
for a composite MSID system in the form of
\begin{equation}\label{2}
\begin{split}
(\omega^2-\kappa_s^2-k^2a_s^2+2\pi G|k|\Sigma_0^s)\\
\times[\omega^2-\kappa_g^2-k^2(a_g^2+C_A^2)
+2\pi G|k|\Sigma_0^g]\\
=(2\pi G|k|\Sigma_0^s)(2\pi G|k|\Sigma_0^g)\ .
\end{split}
\end{equation}
As expected, this dispersion relation explicitly shows the mutual 
gravitational coupling between the two SIDs by the term on the 
right-hand side. The first factor on the left-hand side 
%of equation (\ref{2}) 
is for perturbations in the stellar SID approximated as a fluid, 
while the second factor is for coplanar MHD perturbations in the 
gaseous MSID [see expression (\ref{1})]. It should be emphasized 
that while dispersion relation (\ref{2}) is a local one, the 
background physical variables are in rotational MHD equilibrium 
in a consistent manner. In particular, $\Sigma_0^g$ given by 
expression (\ref{121}) contains the information of background 
magnetic field via $C_A^2$.

As noted by Lou \& Shen (2003), the dispersion relation derived here 
for coplanar MHD perturbations in a composite (M)SID approach is 
qualitatively similar to those previously obtained by Jog \& Solomon 
(1984), Elmegreen (1995) and Jog (1996) in spirit, but one major 
distinction is that in the formulation of our composite (M)SID, the 
rotation speeds of the two SIDs $V_s$ and $V_g$ are different in 
general. We thus have inequality $\kappa_s\neq\kappa_g$, while in 
those earlier analyses, $\kappa_s=\kappa_g$ was presumed a priori 
following the assumption of $V_s=V_g$.

This is conceptually related to the phenomenon of asymmetric drift
(e.g. Binney \& Tremaine 1987). Physically, stellar velocity 
dispersions mimic a pressure-like force for the stellar component, 
while the thermal ISM gas and magnetic pressure forces together act 
on the magnetized gas component. In the same gravitational potential 
well determined by the total mass distribution, the difference in 
the stellar pressure-like force and the sum of the gaseous and 
magnetic pressure forces would lead to different $V_s$ and $V_g$ 
and thus the asymmetric drift. The rare situation of $V_s$ and 
$V_g$ being equal may happen under very special circumstances.

\section{AXISYMMETRIC STABILITY ANALYSIS
FOR A COMPOSITE MSID SYSTEM }

We describe below coplanar MHD perturbation analysis for 
the axisymmetric stability of a composite MSID based on 
a low-frequency time-dependent WKBJ approach. Generalizing 
the notations of Shen \& Lou (2003) yet with the effect of 
magnetic field included, we here define
\begin{equation}\label{13}
H_1\equiv\kappa_s^2+k^2a_s^2-2\pi G|k|\Sigma_0^s\ ,
\end{equation}
\begin{equation}\label{14}
H_2\equiv\kappa_g^2+k^2(a_g^2+C_A^2)-2\pi G|k|\Sigma_0^g\ ,
\end{equation}
\begin{equation}\label{15}
G_1\equiv2\pi G|k|\Sigma_0^s\ ,
\end{equation}
\begin{equation}\label{16}
G_2\equiv 2\pi G|k|\Sigma_0^g\ ,
\end{equation}
where $\Sigma_0^s$ and $\Sigma_0^g$ are given by equations (\ref{120}) 
and (\ref{121}). In addition to the appearance of $C_A^2$ in equation 
(\ref{14}), $\Sigma_0^g$ given by expression (\ref{121}) also contains 
the magnetic field effect.
%{\bf Here the two repetitive equations are removed.}
%\begin{equation}\label{17}
%\Sigma_0^s=\frac{a_s^2(1+D_s^2)}{2\pi Gr}\frac{1}{(1+\delta)}\ ,
%\end{equation}
%\begin{equation}\label{18}
%\Sigma_0^g=\frac{a_g^2(1+D_g^2)-C_A^2/2}{2\pi Gr}
%\frac{\delta}{(1+\delta)}\ .
%\end{equation}
Dispersion relation (\ref{2}) 
can be cast into the form of
\begin{equation}\label{31}
\omega^4-(H_1+H_2)\omega^2+(H_1H_2-G_1G_2)=0\ ,
\end{equation}
with the two roots $\omega^2_{+}$
and $\omega^2_{-}$ given by
\begin{equation}\label{32}
\begin{split}
&\omega_{\pm}^2(k)=(H_1+H_2)/2\\
&\qquad\qquad
\pm[(H_1+H_2)^2-4(H_1H_2-G_1G_2)]^{1/2}/2 \ .
\end{split}
\end{equation}
Similar to the proof of Shen \& Lou (2003, 2004a),
the $\omega_{+}^2$ root remains always positive. 
%{\it (This is really an easy proof which is completely
%similar to that of Shen \& Lou 2003)} ({\bf Is this an
%easy proof? Or, a reference to Shen \& Lou (2003, 2004a)?}).
In contrast, $\omega_{-}^2$ may become negative, 
leading to axisymmetric MSID instabilities.
%{\bf Here a repetitive equation is removed.}
%\begin{equation}\label{33}
%\omega_{-}^2(k)=\frac12\{(H_1+H_2)-[(H_1+H_2)^2
%-4(H_1H_2-G_1G_2)]^{1/2}\}.
%\end{equation}
%Therefore this $\omega_{-}^2$ is the key for examining
%the stability property of the composite MSID system.
Substitutions of expressions $H_1$, $H_2$, $G_1$, $G_2$ and
definitions (\ref{120}) and (\ref{121}) and expression (\ref{186})
into equation (\ref{32}) for the minus-sign solution would give
$\omega_{-}^2$ in terms of five dimensionless parameters $D_s^2$,
$K\equiv |k|r$, $\delta$, $\beta$ and $\lambda^2$, namely
\begin{equation}\label{34}
\begin{split}
\omega_{-}^2(k)=\frac{a_s^2}{2r^2}(A_2K^2+A_1K+A_0-\wp^{1/2})\ ,
\end{split}
\end{equation}
where
\begin{equation}\label{35}
A_2\equiv1+1/\beta+\lambda^2/\beta\ ,
\end{equation}
\begin{equation}\label{36}
A_1\equiv-(1+y)\ ,
\end{equation}
\begin{equation}\label{37}
A_0\equiv2+4y+\frac{(\lambda^2-2)}{\beta}\ ,
\end{equation}
\begin{equation}\label{38}
\wp\equiv B_4K^4+B_3K^3+B_2K^2+B_1K+B_0\ ,
\end{equation}
\begin{equation}\label{39}
B_4\equiv[1-(1+\lambda^2)/\beta]^2\ ,
\end{equation}
\begin{equation}\label{40}
B_3\equiv2(1+y)(\delta-1)[1-(1+\lambda^2)/\beta]/(1+\delta)\ ,
\end{equation}
\begin{equation}\label{41}
B_2\equiv
[y^2+2y-3+(8\beta-4-2\lambda^2+2\lambda^4+2\beta\lambda^2)/\beta^2]\ ,
\end{equation}
\begin{equation}\label{42}
B_1\equiv4(1+y)(1-\delta)[1-1/\beta+\lambda^2/(2\beta)]/(1+\delta)\ ,
\end{equation}
\begin{equation}\label{43}
B_0\equiv4[1-1/\beta+\lambda^2/(2\beta)]^2\ ,
\end{equation}
where $y\equiv D_s^2$. The analysis here parallels that of Shen 
\& Lou (2003); the novel magnetic field effect to be explored 
is contained in the dimensionless parameter $\lambda^2$.

As a result of the one-to-one correspondence between $D_s^2$ and 
$D_g^2$ dictated by expression (\ref{186}), it is straightforward 
to derive an equivalent form of expression (\ref{34}) in terms of 
$D_g^2$ instead of $D_s^2$. Mathematical solutions of $D_g^2$ and 
$D_s^2$ become unphysical for either $D_g^2<0$ or $D_s^2<0$ or both 
being negative. One can readily show from condition (\ref{186}) 
that $D_s^2<D_g^2$ for $\beta\geq 1$ (Lou \& Zou 2004). Therefore, 
it suffices to require $D_s^2>0$. In the subsequent analysis, we 
mainly use $D_s^2$ -- the square of the rotational Mach number -- 
to examine the axisymmetric stability property in a composite MSID 
system.

By setting $\lambda^2=0$ for zero magnetic field in expressions 
(\ref{34})$-$(\ref{43}), they all reduce to the corresponding 
expressions for a composite system of two coupled unmagnetized 
SIDs analyzed by Shen \& Lou (2003). For scale-free discs more 
general than SIDs, the reader is referred to the work of Syer 
\& Tremaine (1996), Shen \& Lou (2004a, b) and Shen, Liu \& 
Lou (2005). To derive an effective MHD $Q_M$ parameter for a 
composite disc system of one SID and one gaseous MSID, we must 
determine the value of $K_{min}$ at which $\omega_{-}^2$ reaches 
the minimum value.

\subsection{The $D_s^2-$Criterion in the WKBJ Regime}

Before defining an effective $Q_M$ parameter, we first show
unambiguously the $D_s-$criterion for axisymmetric stability 
and confirm our earlier interpretations for the marginal 
stability curves in a composite MSID system (Lou \& Zou
2004).

According to solution (\ref{34}), $\omega_{-}^2$ is a function of
$K$ and $D_s^2$. By setting $\omega_{-}^2=0$ and assigning values of
parameters $\delta$, $\beta$ and $\lambda^2$ in solution (\ref{34}), we 
end up with an equation for $D_s^2$ and $K$. Contours of $\omega^2_{-}$ 
in $D_s^2$ and $K$ with various combinations of $\delta$, $\beta$ and 
$\lambda^2$ are given below to compare with our marginal stability 
curves obtained earlier (Lou \& Zou 2004; see also Shen \& Lou 2003, 
2004a, b, Shen, Liu \& Lou 2005 and Lou \& Wu 2005).

%{\bf Please discuss the reason for plotting $D_s^2$ only or give
%a %reference (Lou \& Zou 2004?) on this point.} {\it (Relevant
%discussion has been added in the above subsection)}
As an example of illustration, we set $|m|=0$, $\delta=0.2$, 
$\beta=1.5$ and $\lambda^2=1$ and determine numerically 
contour curves of $\omega^2_{-}$ in terms of $D_s^2$ 
and $K$ as displayed in Fig. \ref{f1}.
%{\bf Please modify the figure for $\omega^2_{-}$ and include $m=0$!! } 
Physically, the two regions labelled $\omega_{-}^2<0$ 
in the lower-left and upper-right corners are unstable, 
while the region labelled by $\omega_{-}^2>0$ is stable 
against axisymmetric coplanar MHD perturbations.
%Therefore from such contour plot of $D_s^2$ versus $K$, one 
%can numerically determine the specific range of stellar 
%rotation parameter $D_s^2$ in which the composite MSID system 
%is stable or unstable against axisymmetric perturbations.
For comparison, we show the global marginal stability 
curve in a composite MSID system with the same parameter values in 
Fig. \ref{f2} (figure 11 of Lou \& Zou 2004), where $\alpha$ is a 
dimensionless effective radial wavenumber (see Shu et al. 2000; Lou 
2002; Lou \& Shen 2003; Lou \& Zou 2004). In the WKBJ approximation 
of large $K$ and $\alpha$, the two upper-right solid curves in Figs. 
\ref{f1} and \ref{f2} show good mutual correspondence for the ring 
fragmentation instability. Thus our previous interpretation for the 
global stationary axisymmetric MHD perturbation configuration as the 
marginal stability curve is confirmed by the WKBJ analysis here. 
In comparison, the correspondence in the small $K$ regime is 
qualitative with apparent deviations; the WKBJ approximation works 
better for a local analysis, while the results of Fig. \ref{f2} are 
global and exact without the WKBJ approximation. It is clear that 
this comparison reveals the physical nature of the demarcation 
curves as the axisymmetric stability boundaries.

\begin{figure}
\begin{center}
\includegraphics[angle=0,scale=0.40]{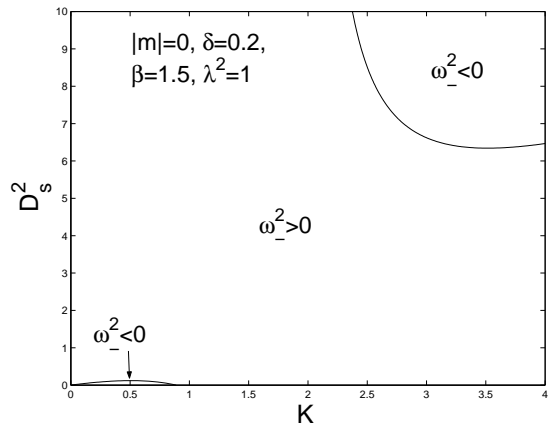}
\caption{\label{f1} A contour plot of $\omega_{-}^2$ as a function
of $K$ and $D_s^2$ with $|m|=0$, $\delta=0.2$, $\beta=1.5$ and
$\lambda^2=1$. The two separated regions labelled $\omega^2_{-}<0$ 
in the lower-left and upper-right corners are unstable. The two 
solid curves mark $\omega^2_{-}=0$.
%{\bf Please label $\omega^2_{-}$ explicitly! Please also mark
%$\omega^2_{-}=0$ in the Figure. }{\it $\omega_{-}^2$ has been
%explicitly labelled in all relevant figures. I do not mark
%$\omega_{-}^2=0$ because I think it would confusing to label both
%$\omega_{-}^2=0$ and $\omega_{-}^2<0$ for the collapse region.} 
}
\end{center}
\end{figure}

\begin{figure}
\begin{center}
\includegraphics[angle=0,scale=0.40]{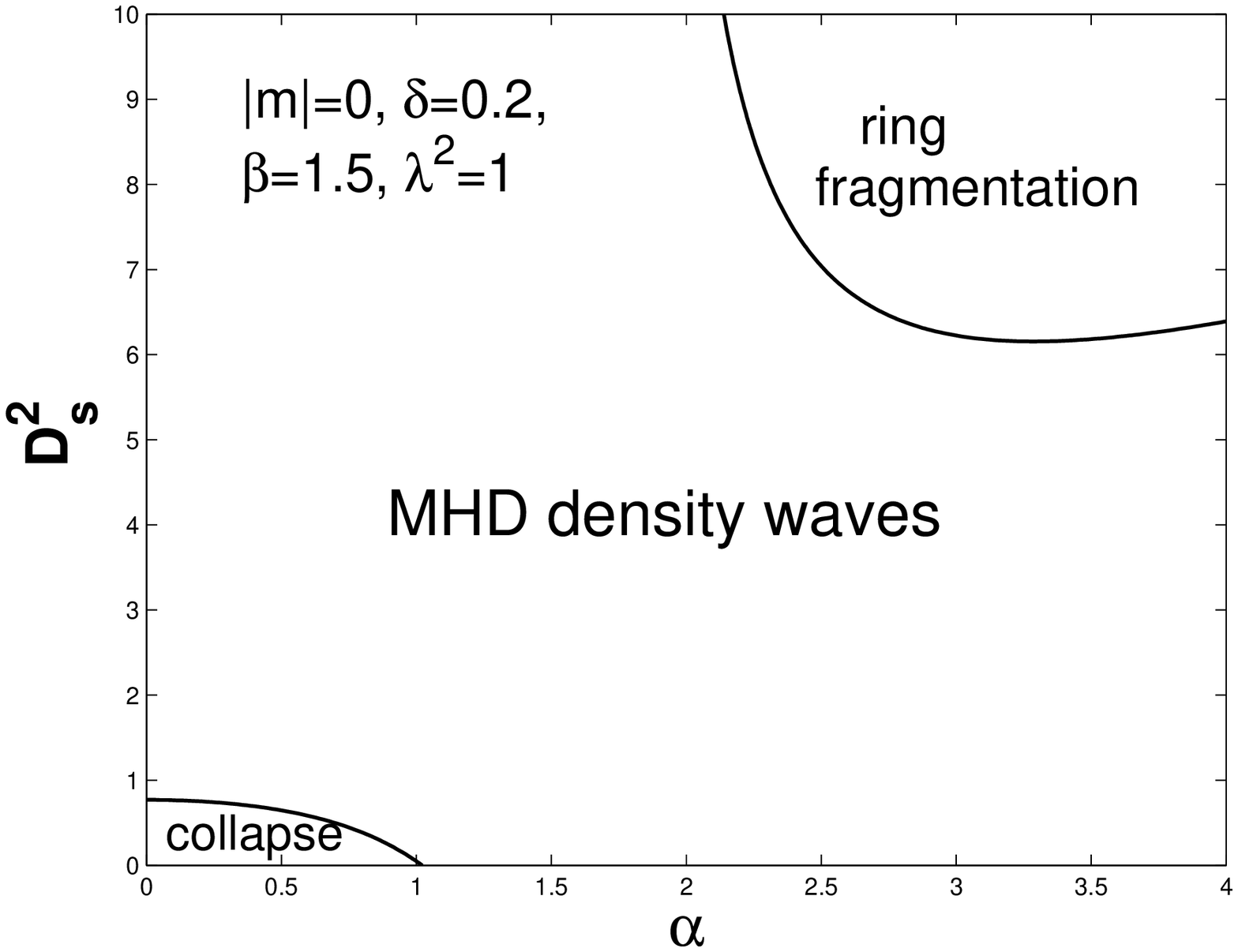}
\caption{\label{f2} The global marginal stability curve of 
$D_s^2$ versus effective dimensionless radial wavenumber 
$\alpha$ for $|m|=0$, $\delta=0.2$, $\beta=1.5$ and 
$\lambda^2=1$ [see figure 11 of Lou \& Zou (2004)].}
\end{center}
\end{figure}

\begin{figure}
\begin{center}
\includegraphics[angle=0,scale=0.40]{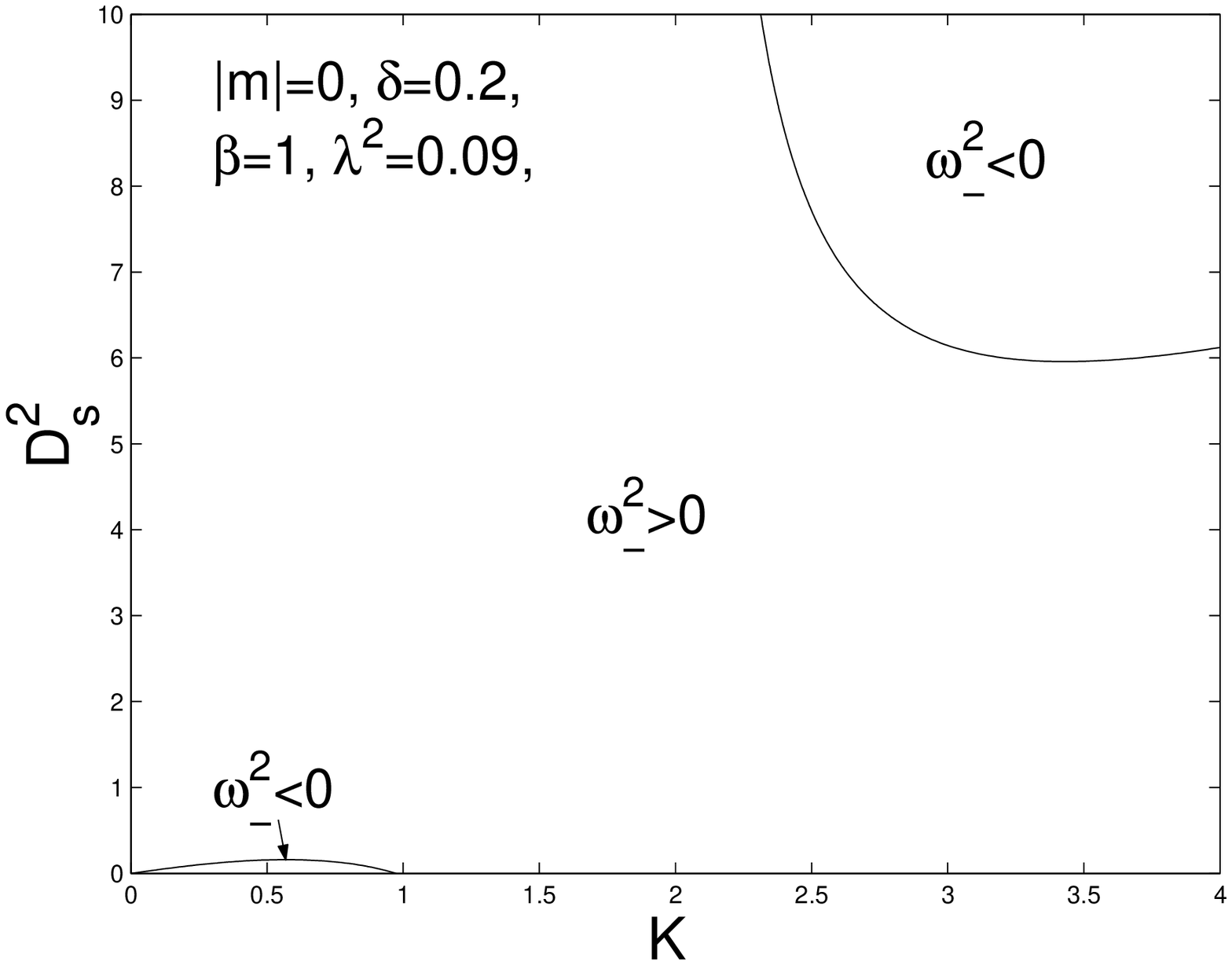}
\caption{\label{f5} A contour plot of $\omega_{-}^2$ as a function of $K$ 
and $D_s^2$ with $|m|=0$, $\delta=0.2$, $\beta=1$ and $\lambda^2=0.09$. 
The two separated domains labelled by $\omega_{-}^2<0$ are unstable. 
The two solid curves mark $\omega_{-}^2=0$. }
\end{center}
\end{figure}

\begin{figure}
\begin{center}
\includegraphics[angle=0,scale=0.40]{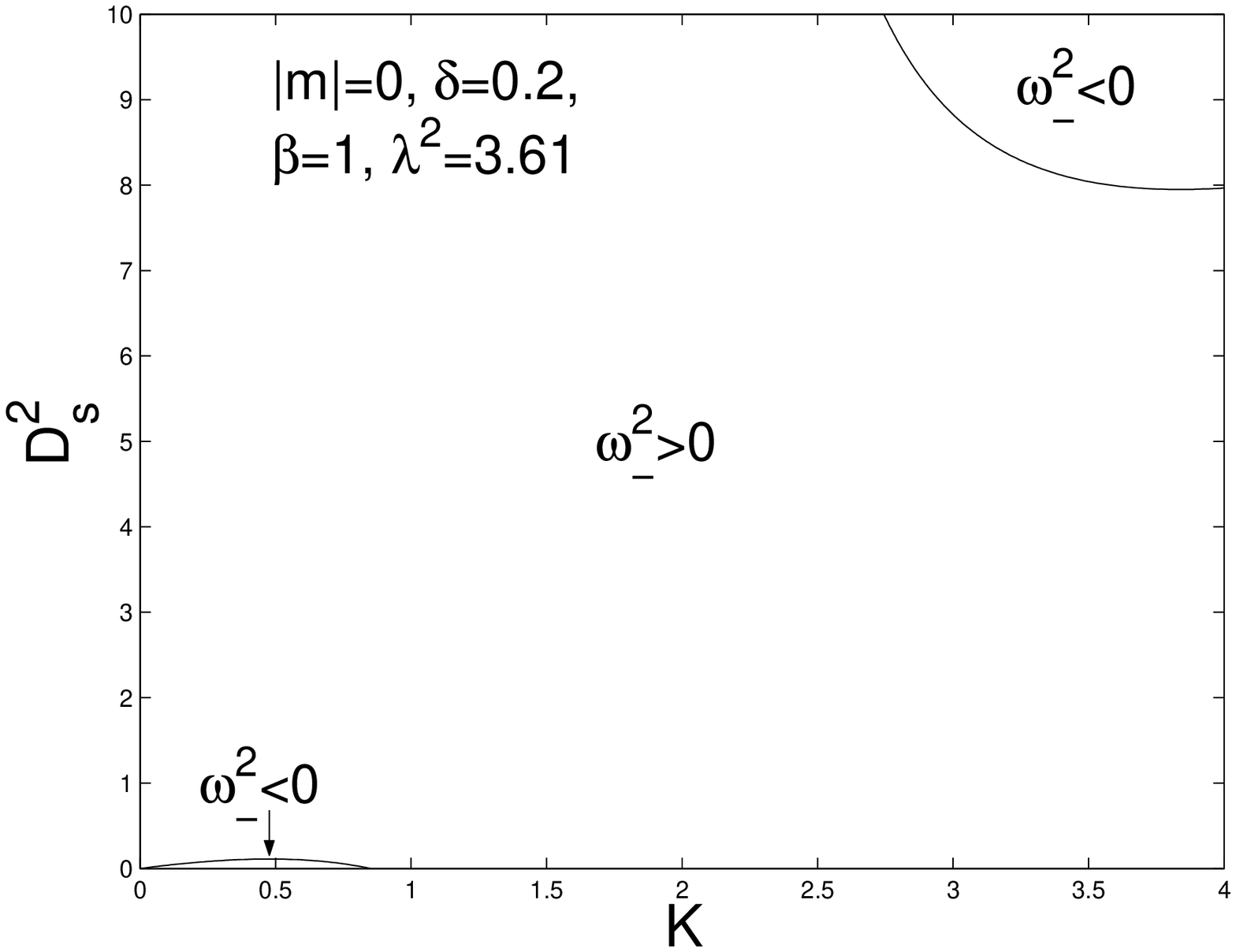}
\caption{\label{f9} A contour plot of $\omega_{-}^2$ as a function
of $K$ and $D_s^2$ with $|m|=0$, $\delta=0.2$, $\beta=1$ and
$\lambda^2=3.61$. The two domains labelled by $\omega_{-}^2<0$ 
are unstable. The two solid curves mark $\omega_{-}^2=0$.  }
\end{center}
\end{figure}

\begin{figure}
\begin{center}
\includegraphics[angle=0,scale=0.40]{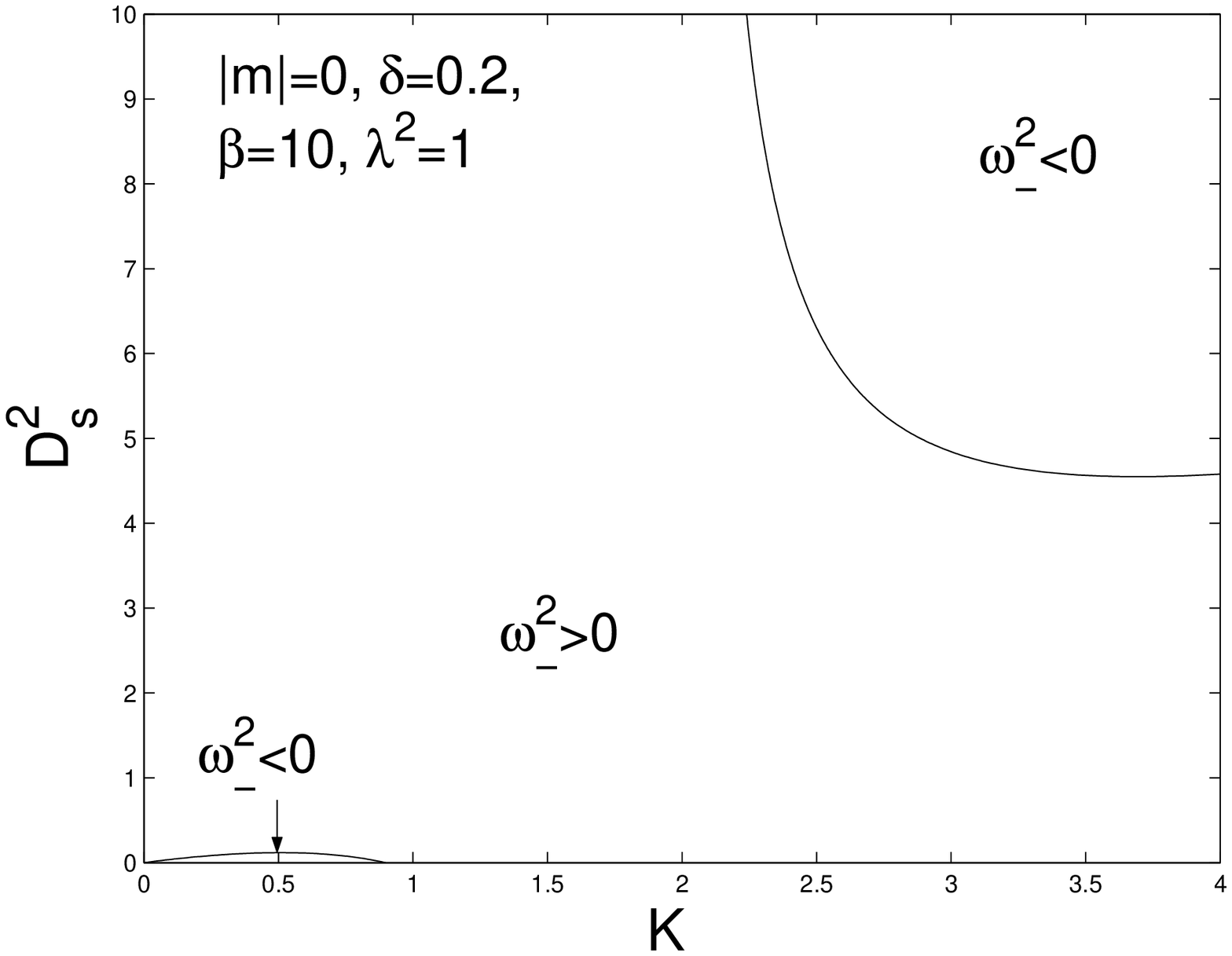}
\caption{\label{f7} A contour plot of $\omega_{-}^2$ as a function
of $K$ and $D_s^2$ with $|m|=0$, $\delta=0.2$, $\beta=10$ and
$\lambda^2=1$. The two domains labelled by $\omega_{-}^2<0$ 
are unstable. The two solid curves mark $\omega_{-}^2=0$. }
\end{center}
\end{figure}

\begin{figure}
\begin{center}
\includegraphics[angle=0,scale=0.40]{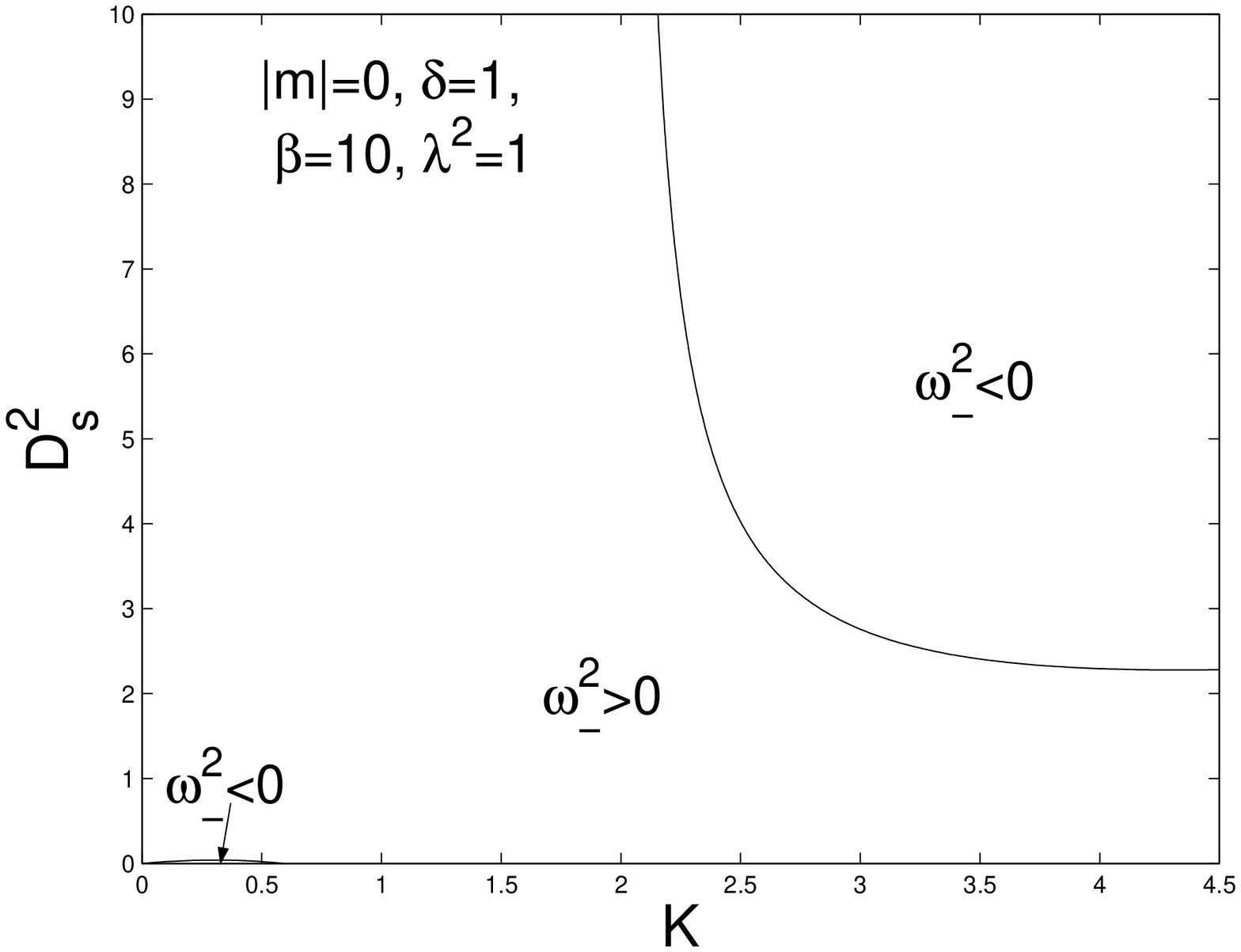}
\caption{\label{f3} A contour plot of $\omega_{-}^2$ as a function 
of $K$ and $D_s^2$ with $|m|=0$, $\delta=1$, $\beta=10$ and 
$\lambda^2=1$. The two regions labelled by $\omega_{-}^2<0$ 
are unstable. The two solid curves mark $\omega_{-}^2=0$.  }
\end{center}
\end{figure}

\begin{figure}
\begin{center}
\includegraphics[angle=0,scale=0.40]{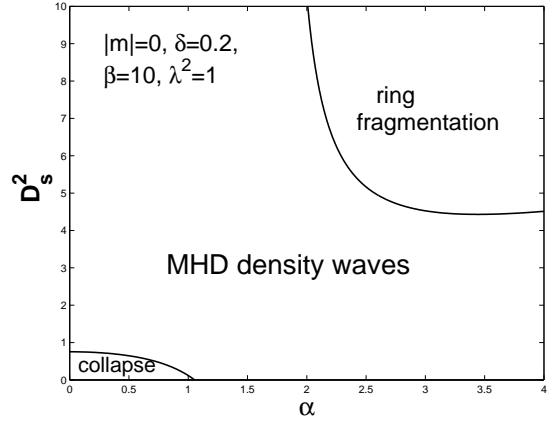}
\caption{\label{f8} The global marginal stability curve of 
$D_s^2$ versus effective dimensionless radial wavenumber 
$\alpha$ for $|m|=0$, $\delta=0.2$, $\beta=10$ and 
$\lambda^2=1$ [see figure 13 of Lou \& Zou (2004)].  }
\end{center}
\end{figure}

By an $\omega_{-}^2$ contour plot for $D_s^2$ versus $K$, the stable 
range of $D_s^2$ in the WKBJ approximation can be readily identified. 
For example, in Fig. \ref{f1} with $|m|=0$, $\delta=0.2$, $\beta=1.5$ 
and $\lambda^2=1$, the composite MSID system has a stable range from 
$D_s^2=0.1205$ at $K=0.5075$ to $D_s^2=6.3428$ at $K=3.5149$. In the 
WKBJ approximation, we explore numerically and show some qualitative 
trends of variations for the marginal stability curves with parameters 
$\delta$, $\beta$ and $\lambda^2$ in Figs. \ref{f5}$-$\ref{f7}. In 
general, the increase of $\delta$ and $\beta$ tends to make a composite 
MSID system more vulnerable to instability (compare Figs. \ref{f1}, 
\ref{f7} and \ref{f3}), while the increase of the magnetic field 
strength $\lambda^2$ expands the stable range and reduces 
the danger of instabilities (compare Figs. \ref{f5} and \ref{f9}). 
%{\bf Please discuss the physical reason of this stability 
%enhancement more explicitly. The background Lorentz force 
%is radially inward and gas surface mass density is reduced. 
%You may discuss this in reference to dispersion relation 
%(\ref{2}) paying attention to radial wavenumber as well.}

The role of coplanar ring magnetic field in stabilizing a composite 
MSID system can be physically understood as follows. According to 
dispersion relation (\ref{2}), we see that, in the MSID factor, 
the gas and magnetic pressure terms are explicitly associated with 
the square of the radial wavenumber $|k|$, while the background 
surface mass density term is linear in $|k|$. For the ring 
fragmentation instability that occurs at relatively large radial 
wavenumbers, the two pressure terms dominate and the increase of 
magnetic pressure tends to enhance the axisymmetric stability of 
a composite MSID system. For the Jeans collapse instability that 
occurs at relatively small wavenumbers, the background gas surface 
mass density term in the MSID factor becomes dominant over the two 
pressure terms. By the rotational MHD radial 
force balance condition (\ref{121}), the background gas surface 
mass density tends to be reduced by the increase of magnetic 
field strength and thus the Jeans collapse instability tends to 
be suppressed. In addition, the right-hand side of dispersion 
relation (\ref{2}) represents the mutual gravitational coupling 
in the presence of coplanar MHD perturbations. A reduction of 
background gas surface mass density will weaken this coupling 
and thus increase the axisymmetric stability.

For a further comparison, we reproduce the global marginal
stability results of figure 13 in Lou \& Zou (2004) here as 
Fig. \ref{f8}, which has the same set of parameters as Fig. 
\ref{f7} for the local WKBJ solution results. We note again 
that for the unstable region (upper right) of large radial 
wavenumber, labelled as the ring fragmentation in Lou \& 
Zou (2004), Figs. \ref{f7} and \ref{f8} show very good 
correspondence as expected. As reference, Table 1 contains 
several lists for the overall stable range of $D_s^2$ with 
$m=0$ and different sets of parameters $\delta$, $\beta$ 
and $\lambda^2$, including the results both from here using the 
WKBJ approximation and from those of Lou \& Zou (2004) for exact 
global MHD perturbation calculations in a composite MSID model.

\begin{table}
\caption{\label{t1}The stable range of $D_s^2$ for axisymmetric
%($|m|=0$) 
coplanar MHD perturbations of any wavelengths for different
values of $\delta$, $\beta$ and $\lambda^2$ in the WKBJ 
approximation. The values in parentheses are those 
determined by Lou \& Zou (2004) for global marginal stability 
curves and are more accurate, especially for describing the 
Jeans collapse regime involving large radial spatial scales.  }
\begin{center}
\begin{tabular}{ccccc}\hline
$\delta$ & $\beta$ & $\lambda^2$ & lower limit of $D_s^2$
& upper limit of $D_s^2$ \\
\hline
0.2 & 1.5 & 1    & 0.1205\ \ (0.7695) & 6.3428\ \ (6.1554)\\
0.2 & 1   & 0.09 & 0.1594\ \ (0.9063) & 5.9573\ \ (5.7561)\\
0.2 & 1   & 3.61 & 0.1111\ \ (0.6783) & 7.9494\ \ (7.7905)\\
0.2 & 10  & 1    & 0.1184\ \ (0.7534) & 4.5499\ \ (4.4310)\\
1   & 10  & 1    & 0.0396\ \ (0.4063) & 2.2787\ \ (2.2434)\\
\hline
\end{tabular}
\end{center}
\end{table}

As shown above, axisymmetric stability properties of a composite 
MSID system can be qualitatively understood using the WKBJ analysis. 
Nevertheless, the WKBJ approximation gradually becomes inaccurate in 
dealing with smaller radial wavenumbers of the Jeans collapse regime. 
Once the physical interpretation has been firmly established, the 
exact global marginal perturbation solution procedure of Lou \& Zou 
(2004) should be adopted to identify the relevant stable range of 
$D_s^2$ in a composite (M)SID system.

\subsection{MHD Generalizations of Effective 
$Q$ Parameters in a Composite MSID System}

For the axisymmetric stability of a single disc, the familiar 
$Q$ parameter (Safronov 1960; Toomre 1964) provides a simple 
local criterion $Q>1$. For a single magnetized disc with a coplanar magnetic 
field, the MHD generalization of this local criterion for the axisymmetric 
stability becomes $Q_M>1$ (Fan \& Lou 1996; Lou \& Fan 1998a). For a 
composite system of two gravitationally coupled discs without magnetic 
field, it would be natural to seek an extension of such a local criterion 
for the axisymmetric stability. Such a local criterion does exist but to 
find a relatively simple analytical expression is not as easy (Elmegreen 
1995; Jog 1996; Shen \& Lou 2003). It should be emphasized that in the 
analyses of both Elmegreen (1995) and Jog (1996), the local background 
variables are prescribed a priori so that the two disc rotation speeds 
and thus the two epicyclic frequencies are the same. In contrast, in the 
analyses of Lou \& Shen (2003) and Shen \& Lou (2003), the equilibrium 
background variables are determined in a dynamically consistent manner so 
that the two SID rotation speeds and thus the two epicyclic frequencies 
are allowed to be different in general. In reference to the local criteria 
of Elmegreen (1995) and Jog (1996), the global $D_s-$criterion of Lou \& 
Shen (2003) for the axisymmetric stability in a composite SID system is 
more straightforward and precise (Shen \& Lou 2003). The main thrust of 
this section is to discuss and establish the global $D_s-$criterion for 
the axisymmetric stability in a composite MSID system with a coplanar 
magnetic field through comparisons (Lou \& Zou 2004).

\subsubsection{MHD Extension of $Q_{E'}$ Parameter of Elmegreen}

Parallelling the procedures of Elmegreen (1995) and of Shen 
\& Lou (2003), we may define an effective $Q_{E}$ parameter 
as the MHD extension of the $Q_{E'}$ parameter of Elmegreen 
for a composite system of MSIDs. From equation (\ref{34}), 
the minimum of $\omega_{-}^2$ is given by
\begin{equation}\label{44}
\begin{split}
\omega_{-min}^2=\frac{a_s^2}{2r^2}
[A_2K_{min}^2+A_1K_{min}+A_0-\wp^{1/2}]\\
=\frac{a_s^2}{2r^2}(\wp^{1/2}-A_2K_{min}^2-A_1K_{min})(Q_{E}^2-1)\ ,
\end{split}
\end{equation}
where
\begin{equation}\label{45}
Q_{E}^2\equiv\frac{A_0}{(\wp^{1/2}-A_2K_{min}^2-A_1K_{min})}\ ,
\end{equation}
and $\wp$ defined by equation (\ref{38}) takes on the value at $K_{min}$. 
Although the forms of these mathematical expressions are strikingly 
similar to those of Elmegreen (1995), all relevant coefficients and 
variables contain the effect of magnetic field througn $\lambda^2$. 
By definition (\ref{45}) of $Q_{E}^2$ above, it is clear that for 
$Q_{E}^2>1$, the minimum of $\omega_{-}^2>0$ and thus the composite 
MSID system would be stable against axisymmetric coplanar MHD 
perturbations of arbitrary radial wavelengths. This generalized 
MHD parameter $Q_{E}^2$
%, thus defined and used in the criterion for axisymmetric stability, 
corresponds to the stable range of $D_s^2$ where $D_s$ is the 
rotational Mach number of the stellar SID modelled as a fluid.

The formidable appearance of $\omega_{-}^2$ given by equation (\ref{44}) 
would prevent us from deriving a straightforward analytical expression 
of $K_{min}$. Nonetheless, instead of minimizing $\omega_{-}^2$ directly, 
it is much simpler to determine the critical value $K_c$ for $K$ 
corresponding to the minimum of variable $\mathcal W\equiv\omega_{+}^2
\omega_{-}^2$. By equation (\ref{31}), we have 
\begin{equation}\label{46}
\mathcal W\equiv\omega_{+}^2\omega_{-}^2=H_1H_2-G_1G_2\ .
\end{equation}
For possible extrema of ${\mathcal W}$, the relevant 
cubic equation that $K_c$ should satisfy is
\begin{equation}\label{47}
\frac{d\mathcal W}{dK}=\frac{d(H_1H_2-G_1G_2)}{dK}=0\ ,
\end{equation}
or equivalently 
\begin{equation}\label{48}
dK^3+aK^2+bK+c=0\ ,
\end{equation}
with the four coefficients explicitly defined by
\begin{equation}\label{49}
d\equiv4\, \left(1+{\lambda}^{2}\right)
\left(\delta+1\right),
\end{equation}
\begin{equation}\label{50}
a\equiv-3\,\left(\beta\,\delta+{\lambda}^{2}+1\right)
\left(y+1 \right),
\end{equation}
\begin{equation}\label{51}
b\equiv2\, \left(\delta+1\right)
\left(2\,\beta\,y-2+{\lambda}^{2}+2\,
\beta+2\,y{\lambda}^{2}+2\,y
\right),
\end{equation}
\begin{equation}\label{52}
c\equiv- \left(y+1\right)
\left(2\,\beta\,y+2\,\beta\,y\delta-2
+{\lambda}^{2}+2\,\beta\right).
\end{equation}
For most parameter regimes under consideration, there is only 
one real solution for the cubic equation (\ref{48}). This real 
solution $K_c$ takes the lengthy but straightforward form of
\begin{equation}\label{53}
K_c=(x-q/2)^{1/3}+(-x-q/2)^{1/3}-a/(3d)\ ,
\end{equation}
where $x\equiv(q^2/4+p^3/27)^{1/2}$, $p\equiv (b/d)-(a/d)^2/3$ and 
$q\equiv 2(a/d)^3/27-ab/(3d^2)+c/d$. We then use this $K_c$ to estimate 
$K_{min}$ and to determine the effective MHD $Q_E$ parameter as the  
generalization of $Q_{E'}$. Relevant curves of MHD $Q_E^2$ versus 
$D_s^2$ corresponding to different values of $\delta$, $\beta$ 
and $\lambda^2$ are displayed in Figures \ref{f11}$-$\ref{f13}.
\begin{figure}
\begin{center}
\includegraphics[angle=0,scale=0.40]{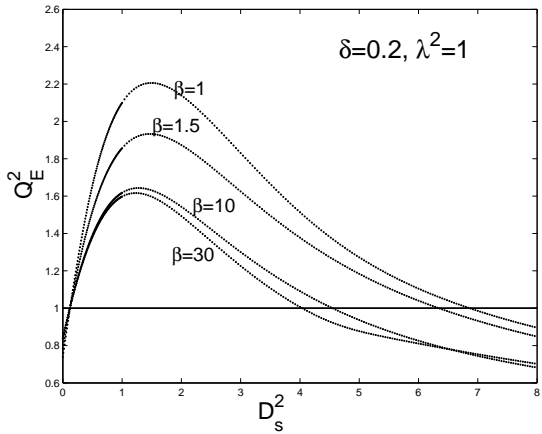}
\caption{\label{f11}Several curves of $Q_E^2$ versus $D_s^2$ 
for different $\beta$ values with specified parameters $|m|=0$,
$\delta=0.2$ and $\lambda^2=1$. For each $Q_E^2$ curve, the two 
intersection points at $Q_E^2=1$ bracket the stable range of 
$D_s^2$. This stable $D_s^2$ range shrinks as $\beta$ increases.
}
\end{center}
\end{figure}

\begin{figure}
\begin{center}
\includegraphics[angle=0,scale=0.40]{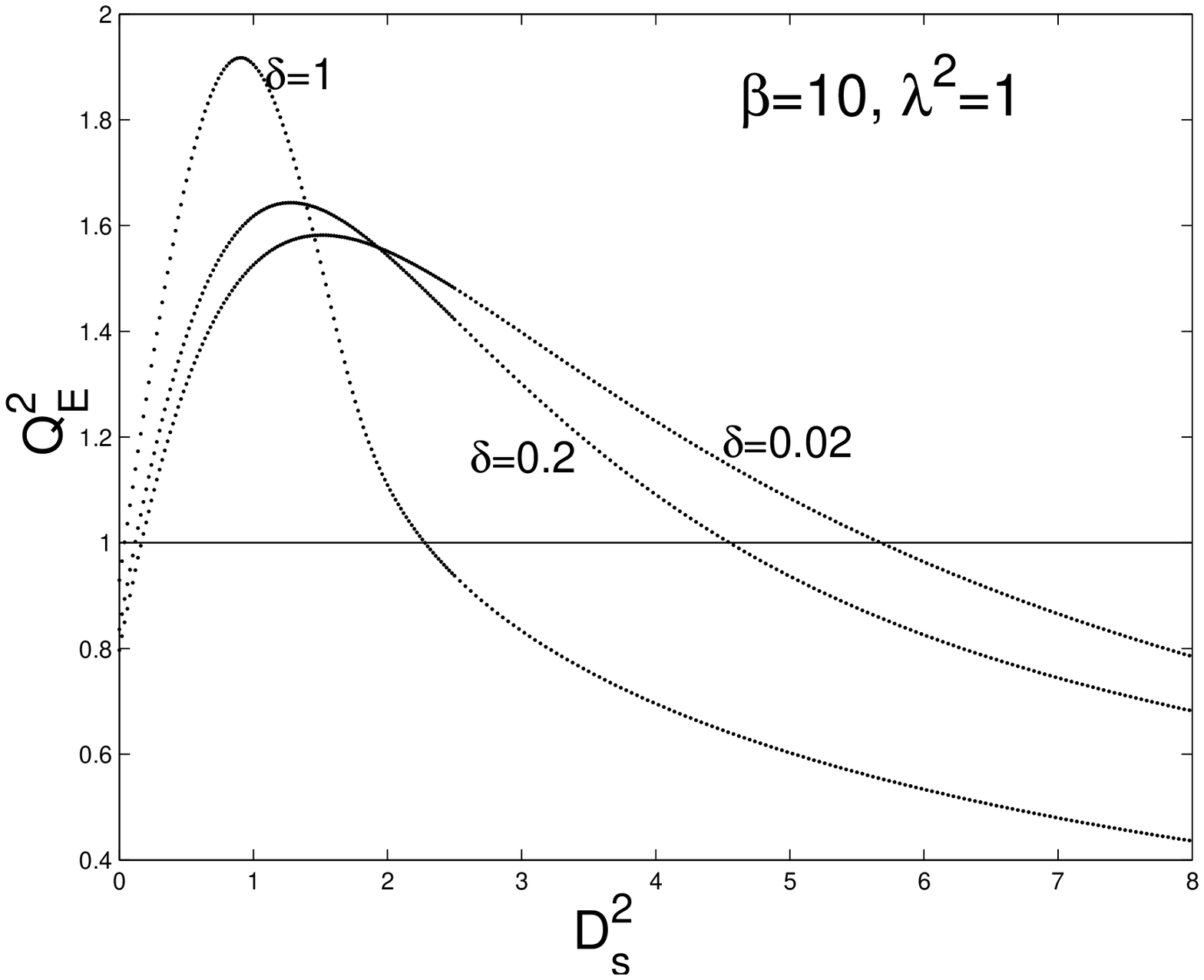}
\caption{\label{f12}Several curves of MHD $Q_E^2$ versus $D_s^2$ 
for different $\delta$ values as indicated. Other relevant 
parameters $|m|=0$, $\beta=10$ and $\lambda^2=1$ are fixed. 
For each MHD $Q_E^2$ curve, the two intersection points at 
$Q_E^2=1$ bracket the stable range of $D_s^2$. This stable 
$D_s^2$ range shrinks as $\delta$ increases. }
\end{center}
\end{figure}

\begin{figure}
\begin{center}
\includegraphics[angle=0,scale=0.40]{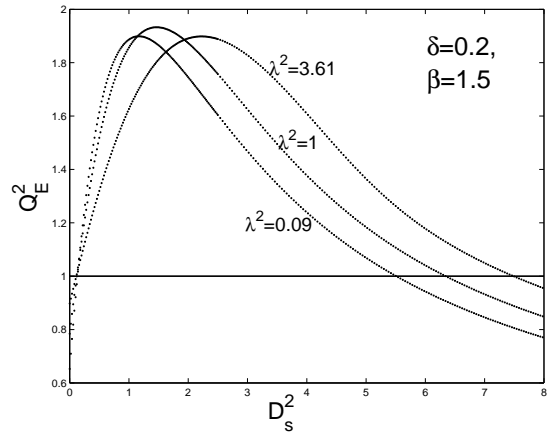}
\caption{\label{f13}Several curves of MHD $Q_E^2$ versus 
$D_s^2$ for different $\lambda^2$ values. Other relevant 
parameters $|m|=0$, $\delta=0.2$ and $\beta=1.5$ are fixed. 
For each curve, the two intersection points at $Q_E^2=1$ 
bracket the stable range of $D_s^2$. This stable $D_s^2$ 
range expands on both ends as $\lambda^2$ increases. }
\end{center}
\end{figure}

By varying parameters $\delta$, $\beta$ and $\lambda^2$, we observe
several trends of variation in the profile of MHD $Q_E^2$ versus 
$D_s^2$ as displayed in Figs. \ref{f11}$-$\ref{f13}. When $\beta$ 
increases with fixed $\delta$ and $\lambda^2$ values, or $\delta$ 
increases with fixed $\beta$ and $\lambda^2$ values, the stable 
range of $D_s^2$ shrinks in general, while the increase of 
$\lambda^2$ tends to expand the stable range of $D_s^2$. Similar 
variation trends have been noticed earlier for the $D_s-$criterion 
and the exact global perturbation solutions in our composite MSID 
model (Lou \& Zou 2004).

\begin{table}
\caption{\label{t2}Approximate stable ranges of $D_s^2$ determined by 
the criterion $Q^2\ge 1$ in a composite system of (M)SIDs. The $D_s^2$ 
values outside the parentheses are given by the MHD $Q_E$ parameter 
generalizing that of Elmegreen (1995), while the $D_s^2$ values inside 
the parentheses are derived from the MHD $Q_J$ parameter generalizing 
that of Jog (1996). These two sets of $D_s^2$ values corresponding to 
MHD $Q_E$ and MHD $Q_J$ parameters are nearly the same and the stable 
$D_s^2$ ranges are very close to the values given by our exact global 
$D_s-$criterion using the MHD perturbation procedure of Lou \& Zou 
(2004; see also Table \ref{t1} here).  }
\begin{center}
\begin{tabular}{ccccc}\hline
$\delta$ & $\beta$ & $\lambda^2$ & lower 
limit of $D_s^2$ & upper limit of $D_s^2$\\
\hline
0.2 & 1.5 & 1    & 0.12\ \ (0.12) & 6.34\ \ (6.34)\\
0.2 & 1   & 0.09 & 0.16\ \ (0.16) & 5.96\ \ (5.96)\\
0.2 & 1   & 3.61 & 0.11\ \ (0.11) & 7.95\ \ (7.95)\\
0.2 & 10  & 1    & 0.12\ \ (0.12) & 4.55\ \ (4.55)\\
1   & 10  & 1    & 0.04\ \ (0.04) & 2.28\ \ (2.28)\\
\hline
\end{tabular}
\end{center}
\end{table}

By definition (\ref{45}) for the MHD $Q_E$ parameter, one determines 
the stable range of $D_s^2$ in the WKBJ approximation as shown in 
Figures \ref{f11}$-$\ref{f13}. For example, given $\delta=0.2$, 
$\beta=1.5$ and $\lambda^2=1$ in Fig. \ref{f11}, we have $Q_E^2>1$ 
for $0.12<D_s^2<6.34$; the composite MSID system is thus stable 
against axisymmetric coplanar MHD perturbations within this range of 
$D_s^2$. More numerical results for the stable ranges of $D_s^2$ for 
different parameters are summarized in Table \ref{t2}. By inspection 
and comparison, the stable $D_s^2$ ranges are almost the same as 
those derived from our $D_s-$criterion as expected (see Table 
\ref{t1}). In comparison with the stable $D_s^2$ ranges derived from 
the exact global axisymmetric perturbation solutions of Lou \& Zou 
(2004), we see again that the results obtained by the two procedures 
have very good correspondence in the larger wavenumber regime, while 
in the Jeans collapse regime of smaller radial wavenumbers, the WKBJ 
approximation leads to apparent deviations. The generalized instability 
criterion characterized by the MHD $Q$ parameter is physically relevant 
to the MHD ring fragmentation instability in a composite disc system of 
gravitationally coupled SID and MSID.

Alternatively, we may follow a similar procedure 
%of Shu et al. (2000), Lou (2002) and described in Appendix B 
of Lou \& Zou (2004) to show the tight
correspondence between the marginal stability curve and the effective
MHD $Q$ parameter. For specific values of parameters $\delta$, $\beta$ 
and $\lambda^2$, we calculate the minimum of the ring fragmentation 
curve for $D_s^2$ using the procedure of Lou \& Zou (2004). Inserting 
the resulting $D_s^2$ into expression (\ref{45}), we readily obtain 
the value of MHD $Q_D^2$ parameter corresponding to the minimum of the 
ring fragmentation curve. For example, given $\delta=0.2$, $\beta=1.5$ 
and $\lambda^2=1$ in figure 11 of Lou \& Zou (2004) (or Fig. \ref{f2} 
here), the minimum of $D_s^2$ in the ring fragmentation curve is about 
6.1554. Using definition (\ref{45}), we obtain the corresponding MHD 
$Q_D^2=1.0216$. More numerical results are summarized in Table \ref{t3}.

\begin{table}
\caption{\label{t3} Numerical values for the minima of $D_s^2$
ring fragmentation curve (Lou \& Zou 2004) and the corresponding 
values of effective MHD $Q$ parameters, including both the MHD 
$Q_E$ parameter generalizing $Q_{E'}$ of Elmegreen (1995) and 
the MHD $Q_J$ parameter generalizing $Q_{J'}$ of Jog (1996).   }
\begin{center}
\begin{tabular}{cccccc}\hline
$\delta$ & $\beta$ & $\lambda^2$
& $(D_s^2)_{min}$ & ${\rm MHD}\hbox{ } Q_E^2$ 
& ${\rm MHD}\hbox{ } Q_J^2$\\
\hline
0.2 & 1.5 & 1    & 6.1554 & 1.0216 & 1.0223\\
0.2 & 1   & 0.09 & 5.7561 & 1.0248 & 1.0247\\
0.2 & 1   & 3.61 & 7.7905 & 1.0144 & 1.0159\\
0.2 & 10  & 1    & 4.4310 & 1.0184 & 1.0188\\
1   & 10  & 1    & 2.2434 & 1.0115 & 1.0102\\
\hline
\end{tabular}
\end{center}
\end{table}

In all these cases, the relevant values of 
MHD $Q_D^2$ are fairly close to unity. Therefore the effective 
MHD $Q$ parameter $Q_D$ is indeed closely relevant to the 
MHD ring fragmentation instability for axisymmetric coplanar 
MHD perturbations in a composite MSID system.

\subsubsection{MHD Extension of $Q_{J'}$ Parameter of Jog}

We have seen that the MHD $Q_E$ parameter generalizing the $Q_{E'}$ of 
Elmegreen (1995) is pertinent to the MHD ring fragmentation perturbation 
in a composite MSID system. However, the determination of the MHD $Q_E$ 
parameter may occasionally become cumbersome when equation (\ref{48}) 
has three real roots of $K$. It would be tedious to identify the 
absolute minimum of $\omega_{-}^2$. Alternative to this procedure, Jog 
(1996) adopted a seminumerical approach to define another effective 
$Q$ parameter for an unmagnetized composite disc system, referred to 
as $Q_{J'}$ here. For comparison, Shen \& Lou (2003) have used a 
generalized procedure to derive the $Q_J$ parameter for a composite 
SID system with a self-consistent background equilibrium allowing two 
different SID rotation speeds but in the absence of magnetic field. 
We shall introduce below the MHD $Q_J$ parameter for a composite 
(M)SID system with a coplanar azimuthal magnetic field and compare 
these results with those of the MHD $Q_E$ parameter.

According to the minus-sign solution (\ref{32}), $\omega_{-}^2$ becomes 
positive and negative depending on whether $H_1H_2-G_1G_2>0$ and $<0$, 
respectively. The critical condition for the axisymmetric stability is 
thus characterized by $H_1H_2-G_1G_2=0$ which can be rearranged into 
the form of
\begin{equation}\label{56}
\frac{2\pi Gk\Sigma_0^g}{[\kappa_g^2+k^2(a_g^2+C_A^2)]}
+\frac{2\pi Gk\Sigma_0^s}{(\kappa_s^2+k^2a_s^2)}=1\ .
\end{equation}
Note that the form of the expression here is strikingly similar to 
that of Jog (1996) with the magnetic terms being both explicit in 
the expression and implicit in $\Sigma^0_g$ through $C_A^2$. With 
magnetic field, we define a new variable $\mathcal F$ by
\begin{equation}\label{54}
\begin{split}
&\mathcal F\equiv\frac{2\pi Gk\Sigma_0^g}
{[\kappa_g^2+k^2(a_g^2+C_A^2)]}+\frac{2\pi Gk\Sigma_0^s}
{(\kappa_s^2+k^2a_s^2)}\\
&\quad
=\frac{K\beta(1+y)\delta/(1+\delta)}
{2[\beta(1+y)-1+\lambda^2/2]+K^2(1+\lambda^2)}\\
&\qquad\qquad\qquad\qquad
+\frac{K(1+y)/(1+\delta)}{(2y+K^2)}\ ,
\end{split}
\end{equation}
where $y\equiv D_s^2$ and $K\equiv|k|r$. We then numerically search 
for $K_{min}$ at which $\omega_{-}^2$ reaches the minimum value. We 
now define an effective MHD $Q_J$ parameter such that
\begin{equation}\label{55}
\begin{split}
&\frac{2}{(1+Q_J^2)}\equiv \mathcal F
=\frac{K_{min}(1+y)/(1+\delta)}{(2y+K_{min}^2)}\\
&\qquad\quad
+\frac{K_{min}\beta(1+y)\delta/(1+\delta)}
{2[\beta(1+y)-1+\lambda^2/2]+K_{min}^2(1+\lambda^2)}\ 
\end{split}
\end{equation}
[see equations (5) and (6) of Jog (1996) and equation (30) of Shen 
\& Lou (2003)]. It follows immediately that $Q_J^2>1$ and $Q_J^2<1$ 
correspond to axisymmetric MHD stability and instability, respectively.
Given $Q_J^2>1$, it indeed follows that $H_1H_2-G_1G_2>0$ for the 
$\omega_{-}^2$ corresponding to $K_{min}$. That is, the minimum of 
$\omega_{-}^2$ is positive for arbitrary $K$ and consequently, the 
composite (M)SID system is stable against axisymmetric coplanar MHD 
disturbances. The procedure of obtaining the MHD $Q_J$ parameter 
can be summarized as follows. For a given set of $\delta$, $\beta$, 
$\lambda^2$ and $D_s^2$, one first determines the value of $K_{min}$ 
numerically using equation (\ref{34}). By inserting this $K_{min}$ 
into equation (\ref{55}), one then obtains the numerical value of 
MHD $Q_J^2$ in a composite MSID system for the given set of $\delta$, 
$\beta$, $\lambda^2$ and $D_s^2$.

\begin{figure}
\begin{center}
\includegraphics[angle=0,scale=0.40]{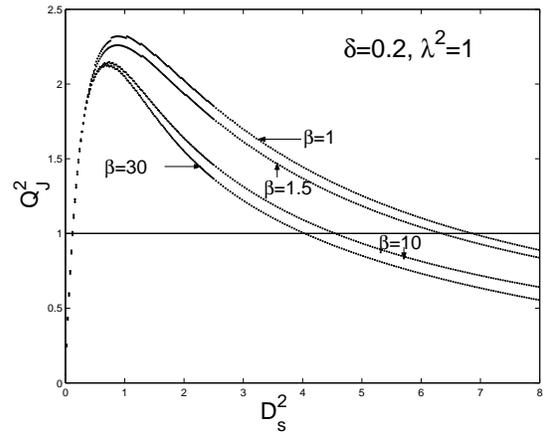}
\caption{\label{f14}Several curves of the MHD 
$Q_J^2$ versus $D_s^2$ for different $\beta$
%$\lambda^2$ 
values. Other relevant parameters $|m|=0$, $\delta=0.2$ 
and $\lambda^2=1$. For each MHD $Q_J^2$ curve, the two 
intersection points at $Q_J^2=1$ bracket the stable 
range of $D_s^2$. This stable $D_s^2$ range shrinks 
as $\beta$ increases. }
\end{center}
\end{figure}

\begin{figure}
\begin{center}
\includegraphics[angle=0,scale=0.40]{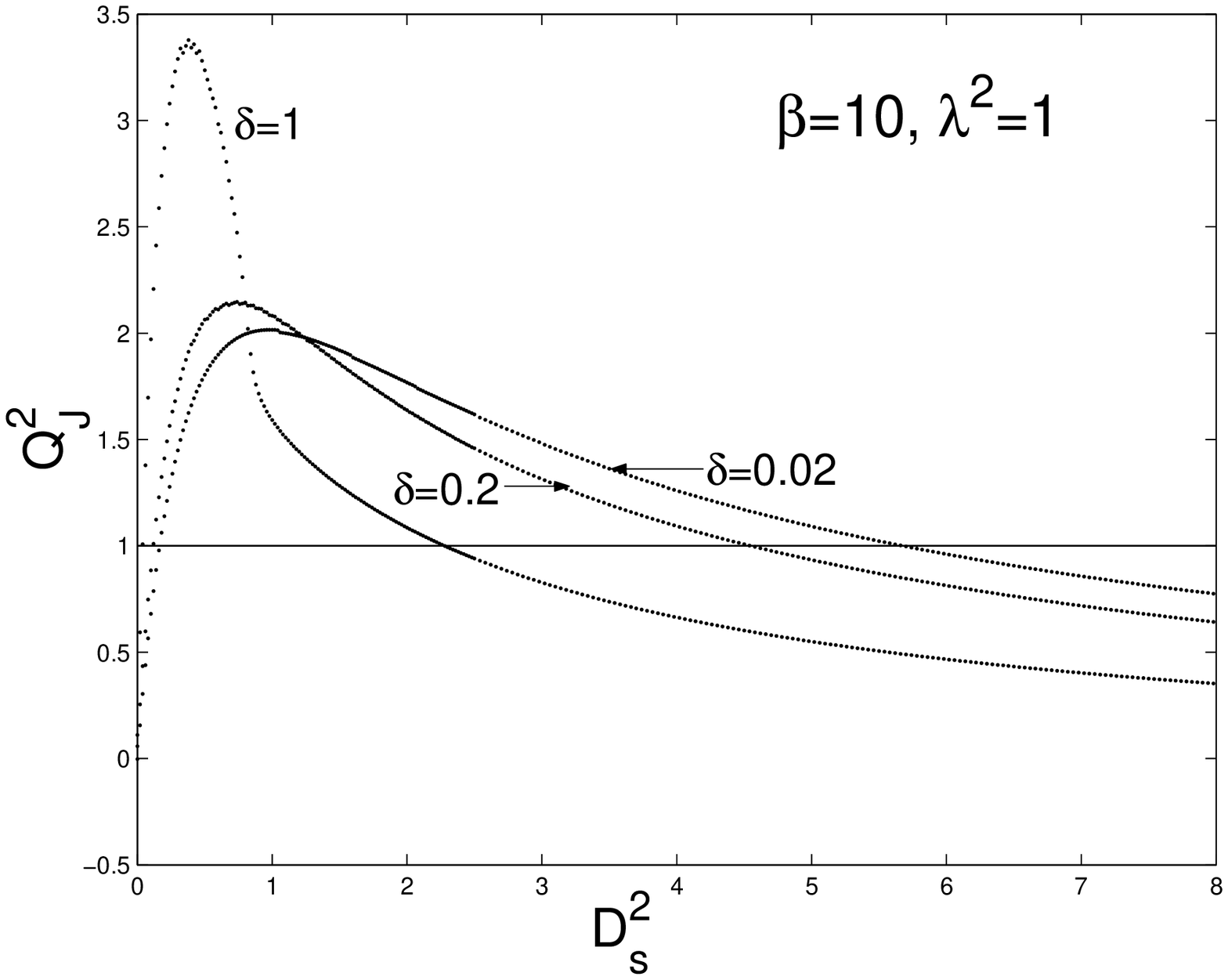}
\caption{\label{f15}Several curves of MHD $Q_J^2$ 
versus $D_s^2$ for different 
%$\lambda^2$ 
$\delta$ values. Other relevant parameters $|m|=0$, $\beta=10$ 
and $\lambda^2=1$ are fixed. For each MHD $Q_J^2$ curve, the two 
intersection points at $Q_J^2=1$ bracket the stable range of 
$D_s^2$. This stable $D_s^2$ range shrinks as $\delta$ increases. }
\end{center}
\end{figure}

\begin{figure}
\begin{center}
\includegraphics[angle=0,scale=0.40]{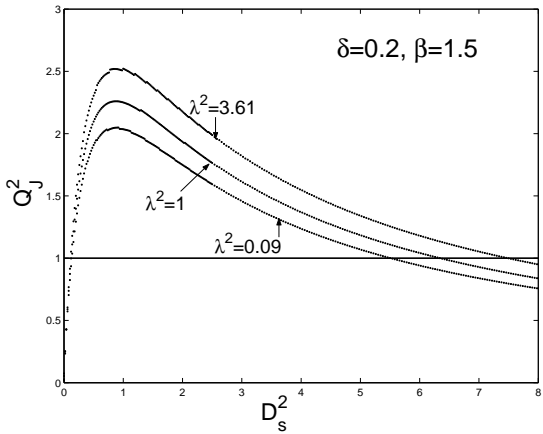}
\caption{\label{f16}Several curves of MHD $Q_J^2$ versus
$D_s^2$ for different $\lambda^2$ values. Other relevant
parameters $|m|=0$, $\delta=0.2$ and $\beta=1.5$ are 
fixed. For each $Q_J^2$ curve, the two intersection 
points at $Q_J^2=1$ bracket the stable range of $D_s^2$. 
This stable $D_s^2$ range expands as $\lambda^2$ increases.}
\end{center}
\end{figure}

We have explored relevant parameter regimes of interest and revealed 
several qualitative variation trends of MHD $Q^2_J$ parameter as shown 
in Figures \ref{f14}$-$\ref{f16}. Similar to the MHD $Q^2_E$ parameter, 
the range of $D_s^2$ for MHD $Q_J^2>1$ corresponds to the axisymmetric 
stability of a composite MSID system. Corresponding to MHD $Q_J^2>1$,
we have calculated stable ranges of $D_s^2$ given several sets of 
$\delta$, $\beta$ and $\lambda^2$ and the detailed results are 
summarized in Table \ref{t2} along with those corresponding to the MHD 
$Q^2_E$ parameter. It is apparent that the stable range of MHD $D_s^2$ 
given by MHD $Q_J^2\ge 1$ is almost the same as that given by MHD 
$Q_E^2\ge 1$. Meanwhile, in Figures \ref{f14}$-$\ref{f16}, we reveal 
the same variation trends as those obtained by MHD $Q_E^2$ and by exact 
global MHD perturbation procedure of Lou \& Zou (2004). When $\beta$ 
increases with fixed $\delta$ and $\lambda^2$ values (see Fig. \ref{f14}), 
or $\delta$ increases with fixed $\beta$ and $\lambda^2$ values (see 
Fig. \ref{f15}), the stable range of $D_s^2$ tends to shrink, while 
the increase of $\lambda^2$ tends to expand the stable range of $D_s^2$ 
(see Fig. \ref{f16}).

Once again, we follow the similar procedure 
%of Shu et al. (2000), Lou (2002) and Appendix B 
of Lou \& Zou (2004) to reveal the close 
relationship between the minimum of the $D_s^2$ marginal stability
curve and the effective MHD $Q$ parameter. For specified values of
parameters $\delta$, $\beta$ and $\lambda^2$, we compute the minimum 
of the ring fragmentation $D_s^2$ curve using the exact global MHD
perturbation procedure of Lou \& Zou (2004), insert the resulting
$D_s^2$ into expression (\ref{55}) and obtain the value of MHD $Q_J^2$ 
parameter corresponding to the minimum of the ring fragmentation curve. 
For example, given $\delta=0.2$, $\beta=1.5$ and $\lambda^2=1$ in 
figure 11 of Lou \& Zou (2004) (or Figure \ref{f2} here), the minimum 
of $D_s^2$ in the ring fragmentation curve is $\sim 6.1554$. Using 
definition (\ref{55}), we obtain the corresponding MHD $Q_J^2=1.02223$. 
More detailed numerical results are summarized in Table \ref{t3} for
reference.

By these numerical experiments, we demonstrate that the values of MHD 
$Q_J^2$ and MHD $Q_E^2$ corresponding to the minima of $D_s^2$ ring 
fragmentation curves are nearly the same, with the relevant values 
of MHD $Q_J^2$ being very close to unity. Therefore, the effective 
MHD $Q_J^2$ parameter is also pertinent to the MHD ring fragmentation 
instability for axisymmetric coplanar MHD perturbations in a composite 
MSID system.

In comparison to the determination of MHD $Q_E^2$, the search for the 
MHD $Q_J$ value requires numerical explorations for each given $D_s^2$. 
The major advantage is that the definition (\ref{55}) for MHD $Q_J^2$ 
remains valid for the entire parameter regime and avoids improper 
situations of unusual sets of $\delta$, $\beta$ and $\lambda^2$.

\subsection{A Composite Partial MSID System }

In most disc or spiral galaxies, there are overwhelming observational 
evidence for the existence of massive dark matter haloes in general. 
To include the large-scale gravitational effect of a massive dark 
matter halo, we add a gravitational potential $\Phi$ term associated 
with the dark matter halo in our basic MHD equations (\ref{22}), 
(\ref{23}), (\ref{25}) and (\ref{26}), where $\Phi$ is presumed to be 
axisymmetric for simplicity and for the lack of information. Based on 
$N-$body numerical simulations for galaxy formation, typical velocity 
dispersions of dark matter `particles' are fairly high (more than a few 
hundred kilometers per second). Hence, another major simplification of
our analysis is to ignore perturbation responses of the massive dark 
matter halo to coplanar MHD perturbations in a composite MSID system 
(e.g., Syer \& Tremaine 1996; Shu et al. 2000; Lou 2002; Lou \& Fan 
2002; Lou \& Shen 2003; Shen \& Lou 2004a, b). As before, we introduce a 
dimensionless ratio $F\equiv\varphi/(\varphi+\Phi)$ for the fraction of 
the disc potential relative to the total potential all in a background 
equilibrium state (e.g., Syer \& Tremaine 1996; Shu et al. 2000; Lou 
2002; Lou \& Shen 2003; Lou \& Zou 2004). The background rotational 
MHD equilibrium of a composite MSID is thus strongly modified by this 
additional $\Phi$ term. As before, we write $\Omega_s=a_sD_s/r$, 
$\Omega_g=a_gD_g/r$, $\kappa_s=\sqrt2\Omega_s$ and 
$\kappa_g=\sqrt2\Omega_g$. It follows from the radial 
force balance in the (M)SID system that the background 
surface mass densities now become
\begin{equation}\label{57}
\Sigma_0^s=F\frac{a_s^2(1+D_s^2)}{2\pi Gr(1+\delta)}\ ,
\end{equation}
\begin{equation}\label{58}
\Sigma_0^g=F\frac{ [a_g^2(1+D_g^2)-C_A^2/2]
\delta }{2\pi Gr(1+\delta)}\ ,
\end{equation}
where $0\leq F< 1$ for a partial composite MSID and $F=1$ for a full 
composite MSID that has been studied in detail in the previous subsections.
Performing the standard MHD perturbation analysis, we linearize
dependent physical variables but ignore dynamical feedbacks from the
massive dark matter halo to coplanar MHD perturbations in the MSID
system. In the WKBJ approximation, it is then straightforward to
derive a strikingly similar dispersion relation in the form of equation 
(\ref{2}) but with modified background equilibrium properties (\ref{57}) 
and (\ref{58}). Following the same procedure of $D_s-$criterion analysis 
described in section 4.1, we obtain the $\omega_{-}^2$ contour plot as a 
function of stellar rotational Mach number $D_s^2$ and radial wavenumber 
$K\equiv|k|r$ with the potential ratio $F$ as an additional parameter. 
Typical results are displayed in Fig. \ref{f19} as an example of illustration.

%\begin{figure}
%\begin{center}
%\includegraphics[angle=0,scale=0.45]{F=1.eps}
%\caption{\label{f17}An $\omega_{-}^2$ contour plot as a
%function of $K$ and $D_s^2$ with $m=0$, $\delta=0.2$,
%$\beta=1.5$, $\lambda^2=1$ and $F=1$. The two regions
%labelled $\omega_{-}^2<0$ are unstable.}
%\end{center}
%\end{figure}

\begin{figure}
\begin{center}
\includegraphics[angle=0,scale=0.40]{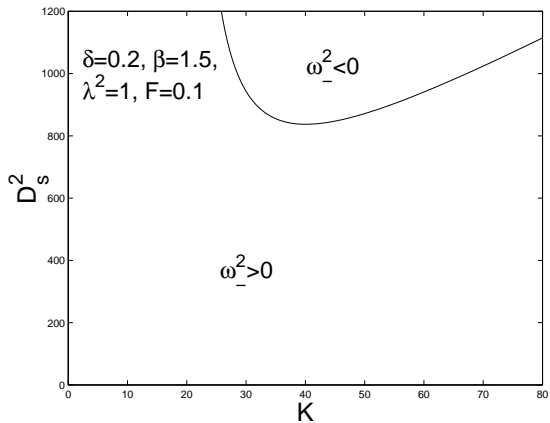}
\caption{\label{f19}An $\omega_{-}^2$ contour plot as a function
of $K$ and $D_s^2$ with $m=0$, $\delta=0.2$, $\beta=1.5$,
$\lambda^2=1$ and $F=0.1$. The region labelled by $\omega_{-}^2<0$
is unstable and the Jeans collapse regime disappears completely. 
In comparison with Fig. \ref{f1}, it is apparent that the stable
range of $D_s^2$ is enlarged as $F$ decreases.    }
\end{center}
\end{figure}
The example of Fig. \ref{f1} with $F=1$ corresponds to a 
full composite MSID system where no background dark matter 
halo is involved, as has been studied in subsection 3.1.
As $F$ becomes less than $1$ corresponding to an increase of the 
potential fraction of the massive dark matter halo, the stable 
range of $D_s^2$ becomes enlarged as clearly shown in Figure 
\ref{f19} for the case of $F=0.1$. From these $\omega_{-}^2<0$ 
contour plots, it is apparent that the introduction of a massive 
dark matter halo tends to stabilize a composite MSID system as 
expected (Ostriker \& Peebles 1973; Binney \& Tremaine 1987; Lou 
\& Shen 2003). For late-type spiral galaxies, one may take $F=0.1$ 
or smaller. Such a composite partial MSID system are stable against 
axisymmetric coplanar MHD disturbances in a wide range of $D_s^2$.

\subsection{Quantitative Estimates and\\
\qquad\  Galactic Applications}

For our numerical examples, the disc mass density ratio parameter
$\delta\equiv\Sigma_0^g/\Sigma_0^s$ has been taken to be 0.2 and 
0.02 (see Fig. \ref{f15}). For late-type spiral galaxies, this ratio 
ranges from 0.05 to 0.15. For relatively young and gas-rich spiral 
galaxies, this ratio $\delta$ can reach 0.2 and even higher. In nearby 
spiral galaxies, the strength of magnetic field is typically inferred 
to be a few to $10\mu G$; by the equipartition argument, magnetic field 
strength may reach a few tens of $\mu G$ in circumnuclear regions and 
within towards the centre (e.g., Lou et al. 2001). We shall take the ratio 
$\lambda\equiv C_A/a_g$ to be of the order of 1. As estimates, we have 
taken $\beta\equiv (a_s/a_g)^2$ to be 1, 1.5, 10, and 30 in Figure \ref{f14}. 
For spiral galaxies, the ratio $a_s/a_g$ can be of the order of or greater 
than 5 or 6. With these estimates in Figs. \ref{f8}$-$\ref{f15}, we see 
clearly that without a massive dark matter halo, a typical disc galaxy 
system with a sufficiently fast rotation (e.g., a $V_s\sim 
150-250\hbox{ km s}^{-1}$) would suffer the ring-fragmentation instability. 
Although the presence of magnetic field offers the stabilizing effect (see 
Fig. \ref{f12}) against the ring-fragmentation instability, the development 
of such instability would be unavoidable for typically inferred galactic 
magnetic field strenghs. By Fig. \ref{f19}, the inclusion of a massive dark 
matter halo holds the key to prevent such ring-fragmentation instability.

In the observational studies of Kennicutt (1989) on global star formation
rates in spiral galaxies, the theoretical rationale is therefore not
sufficiently strong. Firstly, the application of the Toomore stability 
criterion for a single disc is too simple to be physically sensible. 
Secondly, for a composite disc system without magnetic field, the 
generalized criteria (Elmegreen 1995; Jog 1996; Lou \& Fan 1998) for the 
Toomre instability cannot be readily applied to a real galactic disc.
Thirdly, even for a composite disc system with a coplanar magnetic field
which shows clear stabilizing effects, the MHD ring-fragmentation would 
occur for typically inferred parameters of a spiral galaxy without a 
sufficiently massive dark matter halo. Finally, our conclusion is
therefore that in relating Toomre-type instabilities with global star 
formation rates in spiral galaxies, one should conceive new physical 
rationales by incorporating the dynamical interplay between a massive 
dark matter halo and a magnetized composite disc system (Wang \& Silk 
1994; Silk 1997; Lou \& Fan 2002a, b).

\section{SUMMARY AND DISCUSSION}

In this paper, we examined the axisymmetric MHD linear stability 
properties of a composite system consisting of a stellar SID and a 
gaseous MSID coupled by the mutual gravity and a massive dark matter 
halo, using the WKBJ approximation. Our main purpose is to confirm 
the physical interpretation for the global marginal stability curve 
(Lou \& Zou 2004) and to establish the MHD generalization of the $Q$ 
parameter (Safronov 1960; Toomre 1964) for a composite MSID system 
in reference to the earlier work (e.g., Elmegreen 1995; Jog 1996; 
Lou \& Fan 1998a, b; Lou 2002; Lou \& Shen 2003; Lou \& Zou 2004). 
We now summarize the main theoretical results below.

We have recently constructed exact global solutions for stationary coplanar 
MHD perturbations in a composite system of a stellar SID and a gaseous 
MSID for both aligned and unaligned logarithmic spiral cases (Lou \& 
Zou 2004). Lou \& Zou (2004) have extended the analyses of Shu et al. 
(2000) on an isopedically magnetized SID, of Lou (2002) on a single 
coplanarly magnetized SID, and of Lou \& Shen (2003) and Shen \& Lou 
(2003) on an unmagnetized composite SID system. In a broader perspective, 
a composite MSID system is only a special case belonging to a more general 
class of composite scale-free magnetized disc systems (Syer \& Tremaine 
1996; Shen \& Lou 2004a, b; Shen et al. 2005). In analogy of Shu 
et al. (2000), Lou \& Shen (2003) and Shen \& Lou (2003), Lou \& Zou 
(2004) naturally interpreted $D_s^2$ marginal curves for stationary 
axisymmetric coplanar MHD perturbations with radial propagations as the 
marginal stability curves. Based on the low-frequency time-dependent 
WKBJ analysis here (Shen \& Lou 2003, 2004a), we establish the 
physical scenario for the presence of the two unstable regimes referred 
to as the `MHD ring fragmentation instability' and the `MHD Jeans 
collapse' in a composite MSID system. Consequently, it is intuitively 
appealing and physically reliable to apply our exact global 
$D_s-$criterion for a composite MSID system in order to examine its 
axisymmetric stability and obtain the stable range of $D_s^2$, 
where $D_s$ is the rotational Mach number of the stellar SID.

In the WKBJ approximation, we relate our MHD $D_s-$criterion to the 
two effective $Q$ parameters extended to the MHD regime, namely, 
the MHD $Q_E-$criterion generalizing that of Elmegreen (1995) and 
the MHD $Q_J-$criterion generalizing that of Jog (1996). However, our 
procedures differ from those of Elmegreen (1995) and of Jog (1996), 
because our MHD background of rotational equilibrium is dynamically
self-consistent with $\kappa_s\neq\kappa_g$ in general. We show that 
MHD generalizations of both $Q_E-$ and $Q_J-$criteria lead to nearly
the same stable range for $D_s^2$ extended to the MHD realm. This
confirms the close relation between our MHD $D_s-$criterion and the 
MHD $Q_E-$ and $Q_J-$criteria. Complementarily, we show that the values 
of the MHD $Q_E$ and $Q_J$ corresponding to the minima of the $D_s^2$ 
ring fragmentation curves in the exact global MHD perturbation procedure 
of Lou \& Zou (2004) are all close to unity. Our interpretation of the 
axisymmetric marginal stability curve as the demarcation of stable and 
unstable regimes is physically sensible, and the MHD $Q_E$ and $Q_J$ 
parameters are associated with the MHD ring fragmentation instability 
in a composite MSID.

Finally, we have shown the axisymmetric MHD stability property of a 
composite partial MSID by including the gravitational effect from an 
axisymmetric massive dark matter halo. It is apparent that the dark 
matter halo has a strong stabilizing effect for a composite MSID system.

In addition to theoretical interest of disc instabilities for forming 
large-scale galactic structures (n.b., non-axisymmetric ones are not 
studied here), there has been a keen desire to somehow relate such 
instabilities to global star formation rates in disc galaxies and 
their evolution (e.g., Jog \& Solomon 1984a, b; Kennicutt 1989; Wang 
\& Silk 1994; Silk 1997). The overall chain of star formation processes 
from large-scale disc instabilities, to giant molecular clouds, to cloud 
collapses, to clusters of stars, to disc accretion onto individual stars 
and so on is quite complicated and involves many scales of different
orders of magnitudes. Conceptually, large-scale axisymmetric ring 
structures in a disc must be further broken down {\it non-axisymmetrically} 
into smaller pieces in order to initiate this conceived chain of collapses. 
While various stages of this `chain' have been intensively studied separately, 
the ultimate relation or connection between the large-scale axisymmetric 
instabilities and the global star formation rate remains unclear and needs 
to be established (e.g. Elmegreen 1995; Lou \& Bian 2005). 

Observationally, Kennicutt (1989) attempted to infer an empirical 
relation between the $Q$ parameter of the stellar disc alone and 
the global 
star formation rate. Wang \& Silk (1994) pursued a similar idea with 
an estimate of $Q$ parameter for a composite disc system of two fluid 
discs; however, their approximation for $Q$ parameter may be off too 
much under various relevant situations (Lou \& Fan 1998b, 2000a). Should 
this line of reasoning indeed contain an element of truth for addressing 
the issue of the global star formation rate, then the $Q$ parameter
adopted should really correspond to that of a composite disc system with 
a magnetized gas disc component and in the presence of a massive dark 
matter halo. The basic physical reason behind this suggestion is that 
stars form directly in the magnetized gas disc under the joint 
gravitational influence of the dark matter halo, the stellar disc and 
the magnetized gas disc itself. If this line of reasoning does indeed 
make physical sense, then an interesting possibility arises. That is, 
the dark matter halo may play an important role of regulating global 
star formation rates in disc galaxies and thus galactic evolution. For 
example, if the mass of a dark matter halo is very much greater than 
the mass of a composite disc system, then star formation activities 
become weaker. On the other hand, if the dark matter halo is not 
sufficiently massive, then the disc system rapidly evolves into a bar
system. It is also possible that the dark matter halo is only marginal
to maintain a stability of a composite disc. In this case, the global 
star formation activities in the disc system proceed in a regulated 
manner.

Our MHD model analysis in this paper, highly idealized in many ways, 
does contain several requisite elements for establishing an MHD
generalization of the $Q$ parameter and the corresponding criterion
for axisymmetric stability or instability. Observationally, it would 
be extremely interesting to examine the relation between the MHD
$Q_M$ parameter in galactic systems and global star formation rates.
This is not expected to be a trivial exercise given various sources 
of uncertainties.

By presuming that such axisymmetric disc instabilities characterized 
by either $D^2_s$ or $Q_M$ parameters might somehow hint at or connect 
to the global star formation rate, there are a few model problems 
similar to the current one that can be explored further. For example, 
the models of Shen \& Lou (2004b) and Shen, Liu \& Lou (2005) can 
be combined to construct a composite disc system consisting of two
scale-free discs with the gaseous one being coplanarly magnetized in 
the presence of a dark matter halo. Likewise, the work of Lou \& Wu 
(2005) can be generalized to two coupled scale-free disc with the 
gaseous one being isopedically magnetized in the presence of dark 
matter halo (Lou \& Wu 2006 in preparation). The real situation is 
more complicated. We hope these analyses may offer certain hints and 
insights for different aspects of the problem (Lou \& Bai 2005 in 
preparation).

\section*{ACKNOWLEDGEMENTS}
This research has been supported in part by the ASCI Center for 
Astrophysical Thermonuclear Flashes at the University of Chicago 
under Department of Energy contract B341495, by the Special Funds 
for Major State Basic Science Research Projects of China, by the 
Tsinghua Center for Astrophysics, by the Collaborative Research 
Fund from the National Natural Science Foundation of China (NSFC) 
for Young Outstanding Overseas Chinese Scholars (NSFC 10028306) at 
the National Astronomical Observatory, Chinese Academy of Sciences, 
by NSFC grants 10373009 and 10533020 (YQL) at the Tsinghua 
University, and by the special fund 20050003088 and the Yangtze 
Endowment from the Ministry of Education through the Tsinghua 
University. YQL acknowledges supported visits by Theoretical 
Institute for Advanced Research in Astrophysics (TIARA) of 
Academia Sinica and National Tsinghua University in Taiwan. 
The hospitality and support of 
%the Mullard Space Science Laboratory 
%at University College London, U.K., of 
School of Physics and Astronomy, University of St Andrews, Scotland, 
U.K., and of Centre de Physique des Particules de Marseille 
(CPPM/IN2P3/CNRS) et Universit\'e de la M\'editerran\'ee 
Aix-Marseille II, France are also gratefully acknowledged.
Affiliated institutions of YQL share this contribution.

\end{document}